\documentclass{emulateapj}

\usepackage{natbib}
\usepackage{amsmath,amssymb}

\begin{document}

\title{Penumbral fine structure and driving mechanisms of large-scale flows 
  in simulated sunspots}
\shorttitle{Penumbral fine structure and driving mechanism}

\author{M. Rempel}

\shortauthors{Rempel}
 
\affil{High Altitude Observatory,
    NCAR, P.O. Box 3000, Boulder, CO 80307, USA}

\email{rempel@hao.ucar.edu}

\begin{abstract}
We analyze in detail the penumbral structure found in a recent radiative MHD 
simulation. Near $\tau=1$, the simulation produces penumbral fine structure 
consistent with the observationally inferred interlocking comb structure. 
Fast outflows exceeding $8\,\mbox{km\,s}^{-1}$ are present along almost 
horizontal stretches of the magnetic 
field; in the outer half of the penumbra, we see opposite polarity flux 
indicating flux returning beneath the surface. The bulk of the penumbral 
brightness is maintained by small-scale motions turning over on scales shorter 
than the length of a typical penumbral filament. The resulting vertical rms
velocity at $\tau=1$ is about half of that found in the quiet 
Sun. Radial outflows in the sunspot penumbra have two components.  
In the uppermost few $100$ km, fast outflows are driven primarily through
the horizontal component of the Lorentz force, which is confined to narrow
boundary layers beneath $\tau=1$, while the contribution from horizontal 
pressure gradients is reduced in comparison to granulation as a consequence 
of anisotropy. The resulting Evershed flow reaches its peak velocity near 
$\tau=1$ and falls off rapidly with height. Outflows present in deeper layers
result primarily from a preferred ring-like alignment of convection cells 
surrounding the sunspot. These flows reach amplitudes of about $50\%$ of the
convective rms velocity rather independent of depth. A preference for the 
outflow results from a combination of Lorentz force and pressure driving.
While the Evershed flow dominates by velocity amplitude, most of the mass flux 
is present in deeper layers and likely related to a large-scale moat flow.
\end{abstract}

\keywords{convection -- magnetohydrodynamics -- radiative transfer 
-- sunspots -- Sun: surface magnetism}

\received{}
\accepted{}

\maketitle

\section{Introduction}
Since the discovery of the Evershed effect about a century ago 
\citep{Evershed:1909}, the origin of large-scale outflows
in sunspot penumbrae has been a central element in observational
and theoretical studies of sunspots. Over the past decades, advancements
in ground- and space-based observing capabilities have revealed the
stunning fine structure of sunspot penumbrae that is manifest in the
intensity, magnetic field and velocity structure (see, for example, recent 
reviews by \citealt{Solanki:2003,Thomas:Weiss:2004,Thomas:Weiss:2008} and
high resolution observations by \citealt{Scharmer:etal:2002,Langhans:etal:2005,
Rimmele:Marino:2006,Ichimoto:etal:2007,Ichimoto:etal:2007:sc,
Langhans:etal:2007,Scharmer:etal:2007,Rimmele:2008,Franz:Schlichenmaier:2009,
BellotRubio:etal:2010}).
All quantities show in the penumbra a primarily radial filamentary 
structure. Strong horizontal outflows take place in regions with 
almost horizontal field, embedded in a background of more vertical
field -- which has been referred to as "uncombed penumbra"
\citep{Solanki:Montavon:1993} or "interlocking-comb" structure
\citep{Thomas:Weiss:1992}. The connection between the Evershed flow
and the intensity structure is less clear. While earlier
work pointed toward a flow preferentially in the dark component, more
recently \citet{Schlichenmaier:etal:2005} and \citet{Ichimoto:etal:2007}
showed that Evershed flow and intensity variations show a positive 
correlation in the inner and negative correlation in the outer penumbra.
Another controversial aspect is the depth profile of the Evershed flow.
While \citet{Rimmele:1995} and \citet{Stanchfield:etal:1997} found the
Evershed flow in elevated flow channels, more recent work by
\citet{Schlichenmaier:etal:2004,BellotRubio:etal:2006} and 
\citet{Borrero:etal:2008} points toward a flow primarily in the deep 
photosphere that declines with height.

A variety
of simplified models have been proposed to explain the penumbral fine 
structure \citep[e.g.][]{Danielson:1961,Meyer:Schmidt:1968,Galloway:1975,
Thomas:1988,Degenhardt:1989,Degenhardt:1991,Grosser:1991,Wentzel:1992,
Thomas:Montesinos:1993,Montesinos:Thomas:1997,Schlichenmaier:etal:1998a,
Schlichenmaier:etal:1998b,Spruit:Scharmer:2006,Scharmer:Spruit:2006}.
Studies of idealized magnetoconvection in inclined magnetic field
\citep[see, e.g.,][]{Hurlburt:etal:1996,Hurlburt:etal:2000} revealed traveling 
wave-like convection modes, which produce at the surface a combination
of horizontal flow velocities and pattern motions that have been associated with
flow properties observed in penumbrae. Recently substantial progress was made 
in "realistic" numerical simulations that include of the effects of partial 
ionization and radiative transfer. These models were first applied to sections 
of sunspots \citep{Schuessler:Voegler:2006,Heinemann:etal:2007,Rempel:etal:2009,
Kitiashvili:etal:2009} and later to full sunspots \citep{Rempel:etal:Science,
Rempel:2010:IAU}.

Not all of the simplified models listed above contain a self-consistent
description of the Evershed effect, however, such a flow could be added
as an additional degree of freedom to most of them. The models that
include physical processes responsible for driving large-scale outflows 
are based on either stationary or dynamic flux tube models. In the
case of stationary flux tube models \citep{Meyer:Schmidt:1968,Thomas:1988,
Degenhardt:1989,Degenhardt:1991,Thomas:Montesinos:1993,Montesinos:Thomas:1997},
a pressure difference is imposed at the footpoints of the flux tube, which 
leads to siphon flows. In the dynamic flux tube model of 
\citet{Schlichenmaier:etal:1998a,Schlichenmaier:etal:1998b},
fast outflows result from a combination of hot plasma rising at the inner 
footpoint and additional pressure driving that results from radiative 
cooling in the photosphere. 

In the more recent radiative MHD simulations, the penumbral fine
structure is a byproduct of anisotropic overturning convection and the
Evershed flow has been interpreted as the convective flow component
in the direction of the magnetic field \citep{Scharmer:etal:2008}.
A more detailed analysis of \cite{Rempel:etal:Science} concluded that
convection in penumbra and quiet Sun differ primarily in terms of anisotropy
of the velocity field, while typical convective rms velocities and length
scales of energy and mass transport are comparable. The convective
nature of the penumbra is manifest in all radiative 3D simulations to date
but it is still debated in the context of observational constraints. While
there is evidence for overturning convection in some investigations
\citep{Zakharov:etal:2008,Rimmele:2008,Bharti:etal:2010}, many others 
primarily identify the upflow component of the Evershed flow in the inner 
and downflows in the outer penumbra with little evidence for overturning 
convection \citep{BellotRubio:etal:2005,Ichimoto:etal:2007,
Franz:Schlichenmaier:2009,BellotRubio:etal:2010}.

In this paper, we present a detailed analysis of the model presented in
\cite{Rempel:etal:Science} with special focus on the physical origin of 
large-scale outflows. After a brief description of the numerical model 
in Sect. \ref{sect:model}, we focus first on the photospheric appearance 
of the penumbra in Sect \ref{sect:phot}. In Sect. \ref{sect:forcings}
- \ref{sect:simple-model} we present a detailed analysis of the subsurface 
structure responsible for the driving of large-scale outflows. Sect.
\ref{sect:filaments} analyzes the field geometry and connectivity in 
penumbral flow channels and compares our results with findings from
previous models, in particular models based on the flux tube picture.
Sect. \ref{sect:deep-flow} analyzes deeper reaching outflow components beneath
the penumbra that are not directly associated with the Evershed flow. The 
results are summarized and discussed in Sect. \ref{sect:concl}.

\section{Numerical model}
\label{sect:model}
Our investigation is based on the simulation of a pair of opposite polarity 
sunspots described 
in \citet{Rempel:etal:Science}. The simulation in a $98\times 49$ Mm wide
and $6.1$ Mm deep box was designed to study the formation and structure of
penumbrae under a variety of different field strength and inclination
angles. To this end the simulation was initialized with a pair of opposite
polarity sunspots each having a flux of $1.6\cdot 10^{22}$ Mx each, but 
different field strengths of about $3$ and $4$ kG, respectively. In order to 
focus on details of sunspot fine structure a rather high grid resolution of 
$3072\times 1536 \times 384$ ($32$ km horizontal and $16$ km vertical) was 
used at the expense that this simulation could cover only a rather short 
time span.
While the original presentation in \citet{Rempel:etal:Science} was based on
a run of $1.5$ h in high resolution ($3.5$ h total) we have progressed the
simulation in the meantime to $4$ h in high resolution ($6$ h total).
During the extension of the simulation several aspects of the penumbral
structure evolved. During the later stages of the simulation filaments 
became more radially aligned and the mean intensity profile shows
in the inner penumbra a constant value of about $0.7 I_{\odot}$ with a
more steep drop toward the umbra. The overall properties of the
Evershed flow did not show a significant variation in the time frame covered 
by this simulation; we base our detailed analysis of the physical origin of the
flow pattern on the last hour of this simulation run (starting about
$5$ hours after the initialization). 

We also emphasize for clarification that this simulation uses gray radiative
transfer. When we refer in the following discussion to intensity, we mean the
bolometric intensity, if we refer to optical depth, we mean the optical depth
computed with the Rosseland mean opacity.

\section{Photospheric appearance}
\label{sect:phot}
\subsection{Azimuthal averages in photosphere}
\begin{figure}
  \centering 
  \resizebox{\hsize}{!}{\includegraphics{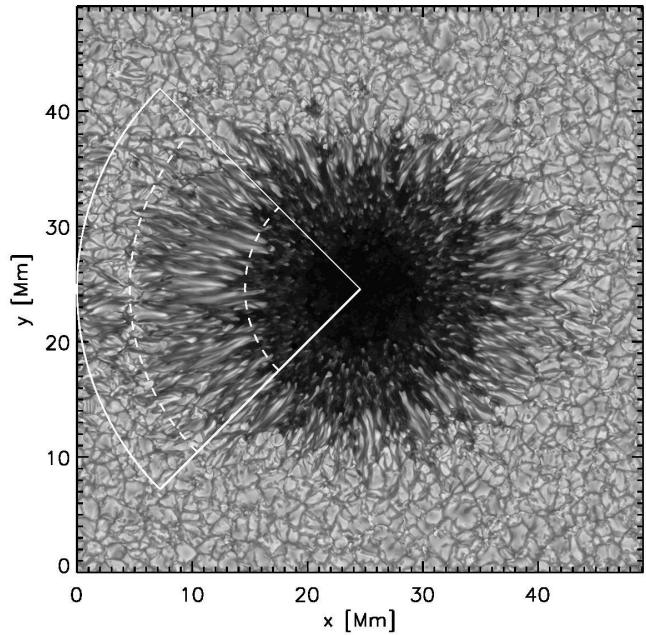}}
  \caption{Intensity image of the sunspot we analyze at about 6 hours 
    after initialization of the simulation run. We show here only one half
    of the horizontal extent of the computational domain, which contains a
    pair of opposite polarity sunspots oriented in the x-direction. The panel 
    is centered on the right 
    sunspot in the simulation domain, which has a stronger magnetic field 
    and is overall more coherent. Our analysis is focused on the 
    penumbral region on the left side. The white lines indicate the region 
    used for azimuthal averages.
  }
  \label{fig:int_region}
\end{figure}

\begin{figure*}
  \centering 
  \resizebox{12.6cm}{!}{\includegraphics{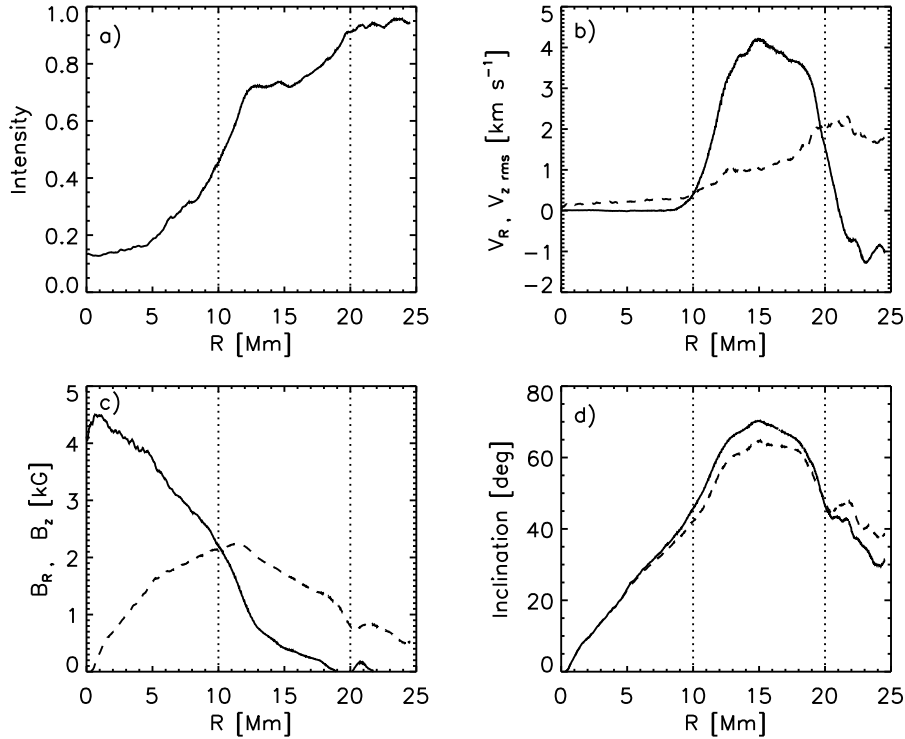}}
  \caption{Azimuthal averages of quantities at the $\tau=1$ level as function
    of the distance from the spot center. The dotted
    vertical lines indicate the radial positions also shown in Fig. 
    \ref{fig:int_region}. Quantities shown are a) intensity relative to quiet 
    sun, b) radial velocity (solid) and vertical rms velocity (dashed),
    c) vertical field strength (solid) and radial field strength 
    (dashed), and d) field inclination with respect to vertical. Here,
    the solid line indicates the average inclination angle, while the dashed
    line indicates the inclination computed from the average field. 
  }
  \label{fig:global_prop}
\end{figure*}

\begin{figure*}
  \centering 
  \resizebox{\hsize}{!}{\includegraphics{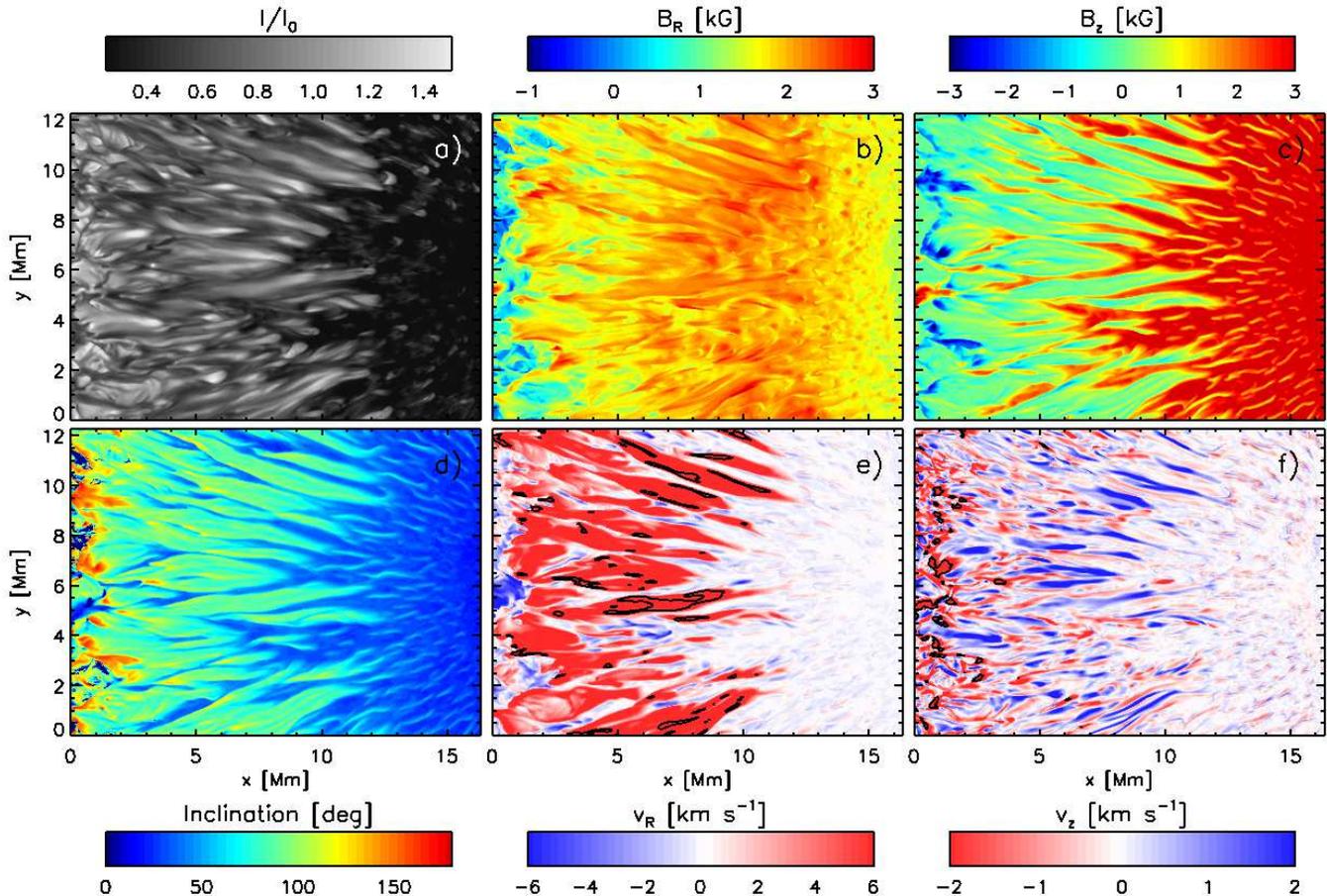}}
  \caption{Sunspot fine structure at the $\tau=1$ level. Quantities shown 
    are a) bolometric intensity, b) radial and c) vertical magnetic field,
    d) field inclination, e) radial and f)
    vertical flow velocities. A field inclination of $0\deg$ corresponds to 
    vertical field with the same polarity as the umbra, $90\deg$ to horizontal 
    and $180\deg$ to vertical field with opposite polarity of the umbra.
    Radial outflows are displayed by red colors, solid contours indicate
    regions with more than $10\,\mbox{km\,s}^{-1}$ outflow velocity. 
    Vertical upflows 
    are displayed by blue colors, solid contours indicate regions with more 
    than $5\,\mbox{km\,s}^{-1}$ downflow velocity.}
  \label{fig:fil_tau1}
\end{figure*}
\begin{figure*}
  \centering 
  \resizebox{12.6cm}{!}{\includegraphics{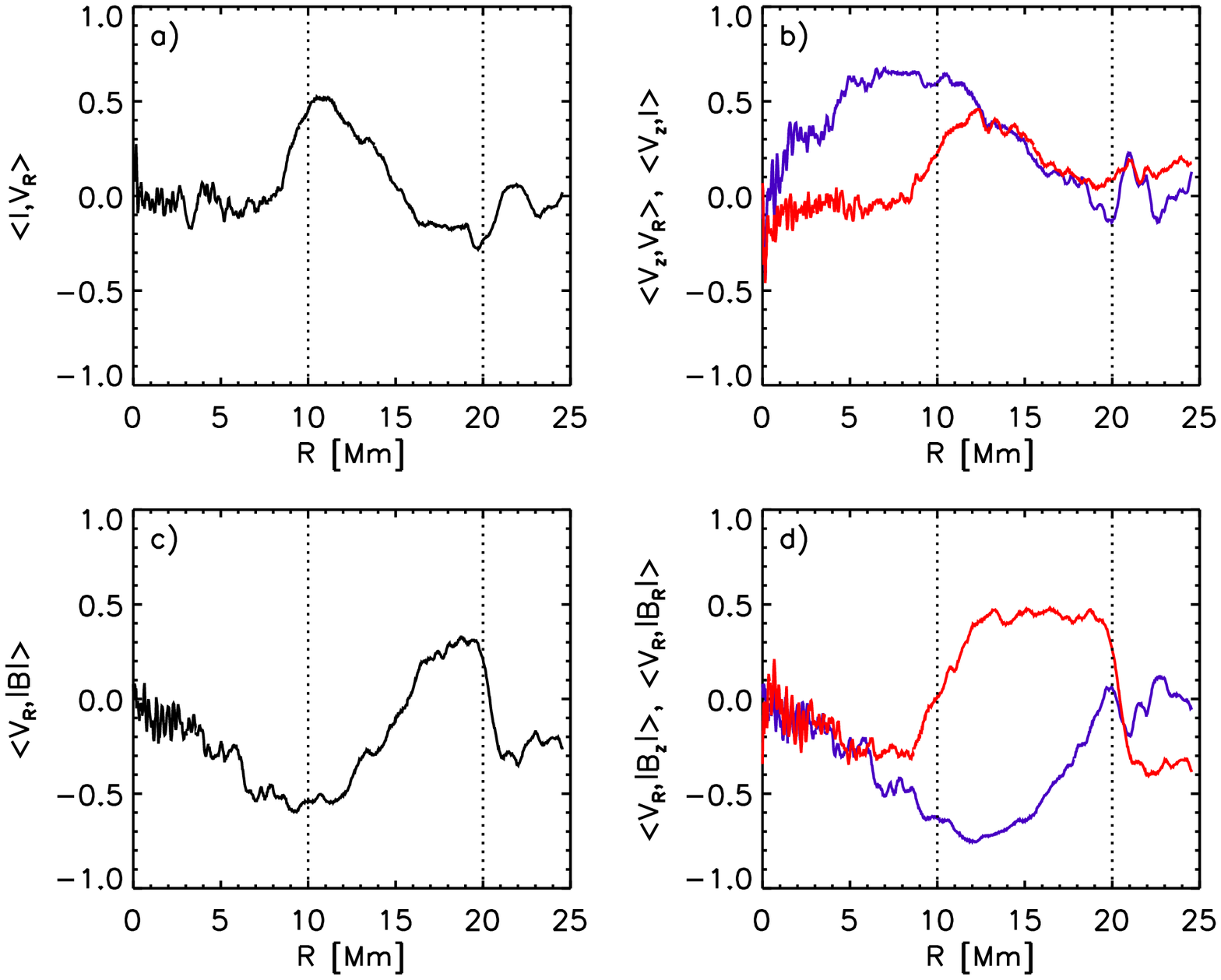}}
  \caption{a) Correlation between intensity and radial velocity 
    ; b) correlation between vertical and radial velocity 
    (blue) as well as vertical velocity and intensity (red); 
    c) correlation 
    between magnetic field strength and radial velocity; d) 
    correlation between vertical field strength and radial velocity (blue) 
    as well as radial field strength and radial velocity (red).
  }
  \label{fig:corr}
\end{figure*}

As described in detail in \citet{Rempel:etal:Science}, the simulation 
domain contains a pair of opposite polarity sunspots, with the most
extended penumbrae found in between the opposite polarity spots along
the horizontal x-direction. The most coherent penumbra is found
in the sunspot with the initially stronger field strength of about $4$
kG \citep[see the spot on the right in Fig. 1 of][]{Rempel:etal:Science}. 
We focus our detailed analysis on the latter, for which an intensity
image is presented in Fig. \ref{fig:int_region}. We are here in particular
interested in the extended penumbra on the left side of the spot for which 
we highlighted the sub-domain used for azimuthal averages in subsequent 
figures.  The dashed lines indicate $R=10$ Mm and $R=20$ Mm from the center
of the spot. Fig. \ref{fig:global_prop} summarizes properties at the
$\tau=1$ level, averaged azimuthally over the $90$ degree wedge shown in
Fig. \ref{fig:int_region} and about $1$ hour in time.
The intensity normalized by the quiet Sun brightness 
$I_{\odot}$ (panel a) shows a sharp increase from umbra toward
penumbra from about 0.15 to 0.7 $I_{\odot}$. In the inner
penumbra, the intensity stays constant on a plateau with about $0.7 I_{\odot}$
and increases then almost linear toward the edge of the penumbra where it 
reaches $0.95 I_{\odot}$ (due to the nearby opposite polarity spot in our 
simulation setup the intensity does not reach $I_{\odot}$). The plateau-like 
intensity profile formed during later stages of this simulation and was not 
present in the results reported earlier by \citet{Rempel:etal:Science}. The 
radial outflow (panel b) starts at about $R=10$ Mm, reaches its peak of 
about $4\,\mbox{km\,s}^{-1}$ near $R=15$ Mm and drops off toward the
outer edge of the penumbra. $R=10$ Mm corresponds to the position at which
the average field inclination (displayed in panel d) angle exceeds $45$ 
degrees, which was already
found by \citet{Rempel:etal:Science} as the critical value for the
onset of large-scale outflows. The position of the peak velocity 
coincides with the position of maximum inclination (about 70 degrees) in 
the middle of the penumbra. The inclination is defined here as 
$\arcsin(B_R/\vert B\vert)$. Due to the strong variation of inclination angle
with azimuth, it makes a difference whether we compute the inclination locally 
and average in azimuth and time later or whether we base the computation on 
the averaged magnetic field presented in panel c). We show in panel d) both, 
the average of the inclination (solid) and the inclination of the average 
field (dashed). The vertical rms velocity (panel b, dashed) increases steadily
throughout the penumbra from a few $100\,\mbox{m\,s}^{-1}$ at $R=10$ Mm 
to about $2\,\mbox{km\,s}^{-1}$
at the outer edge, which is the value corresponding to quiet Sun granulation.
A value of about $1\,\mbox{km\,s}^{-1}$ is required near the inner edge of 
the penumbra ($R=12$ Mm) to maintain the penumbral brightness of 
$0.7 I_{\odot}$. We find in this simulation an approximate relationship of 
the form $I \propto \sqrt{v_{z\,rms}(\tau=1)}$.

\subsection{Filamentation in photosphere}
Fig. \ref{fig:fil_tau1} displays the filamentary fine structure seen
at the $\tau=1$ level in the penumbra. A filamentary structure is present 
in all quantities shown, however the strongest evidence is seen in intensity
(panel a), vertical magnetic field (panel c), inclination (panel d) and 
radial flow velocity (panel e). Penumbral filaments show a strong reduction 
of the vertical magnetic field strength, while the horizontal (radial) field 
component is moderately enhanced (panel b). The combination 
of the two leads to the strong variation of the inclination angle in the 
penumbra. Strong radial outflows (panel e, red color indicates outflows) are 
seen in the 
almost horizontal flow channels, toward the outer end of flow channels 
the inclination angle exceeds $90\deg$, indicating field returning back 
into the convection zone. Radial outflows with more than 
$10\,\mbox{km\,s}^{-1}$ outflow
velocity are indicated by solid contours. Most of the very fast outflows are
found in the inner half of the penumbra, a few of them are associated with
fast downflows in the outer penumbra. The vertical velocity (panel f, blue 
colors indicate upflows) shows strong up and downflows 
everywhere in the penumbra, strong radially aligned upflows are 
preferentially found in the center of penumbral filaments. Solid contours 
indicate regions with more than $5\,\mbox{km\,s}^{-1}$ downflow velocity. 
They are primarily
found near the outer edge of the penumbra. We find downflow speeds of up to
$15\,\mbox{km\,s}^{-1}$ near $\tau=1$. Fast downflows in opposite polarity 
regions have been observed by \citet{Westendorp:etal:2001,DelToro:etal:2001}. 

To clarify the relation between
radial flow velocity, intensity and magnetic field strength in a statistical
sense we present correlation coefficients in Fig. \ref{fig:corr}.
Panel a) displays the correlation between intensity and radial velocity, panel
c) the correlation between field strength and radial
velocity. All correlations are computed based on the fluctuations
of these quantities about their azimuthal mean. Intensity is  
correlated with outflows in the inner penumbra, but weakly anti-correlated 
further outward. The radial outflow is found in regions with reduced field 
strength in the inner, but stronger field in the outer penumbra. Similar 
correlations were found by \citet{Ichimoto:etal:2007} (see Fig. 3 therein)
as well as \citet{Schlichenmaier:etal:2005}. In the panels on the
right (b and d) we present additional correlations, which allow us to make a 
closer connection to the magnetoconvective structure of the penumbra. The
radial dependence of the $I-v_R$ correlation is due to a decorrelation between
vertical and radial velocity in the penumbra (panel b, blue)
combined with a decorrelation of vertical velocity and intensity 
(panel b, red). While the latter remains positive, a sign change 
is present in the former. The physical reason for the decorrelation between 
vertical and radial velocity is evident from the magnetoconvection pattern 
shown in Fig. \ref{fig:fil_tau1}. In the inner penumbra filaments are 
very narrow and the central upflow covers most of the filament, leading to 
large positive correlation between the brighter upflow and radial 
outflow. In the outer penumbra the patches of outflowing material become 
broader and several downflow lanes can be found within these patches, 
resulting in a reduction of the correlation. Toward the outer edge of the 
penumbra stronger downflow patches are present, turning the
correlation weakly negative. Note that the $I-v_z$ correlation stays low
outside $R=20$ Mm due to the proximity of an opposite polarity spot in our
simulation setup. Fig. \ref{fig:corr} panel d)
presents additional correlations between radial velocity and vertical magnetic
field (blue) as well as radial magnetic field strength (red). While the 
former stays 
negative throughout the penumbra, the correlation with the radial field 
strength is positive. In the innermost penumbra the strong reduction
of $\vert B_z\vert$ dominates the picture and leads to an anti-correlation
between $v_R$ and $\vert B\vert$, further out the contribution from the
increased $\vert B_R\vert$ in the flow channels dominates and leads to a 
positive $v_R - \vert B\vert$ correlation. The increase of the inclination
angle found in the flow channels is a consequence of a strong reduction of
$B_z$ to almost zero, while $B_R$ is moderately enhanced in strength. This
asymmetry is due to the fact that $B_R$ benefits from a strong positive
contribution of the induction term $(\vec{B}\cdot\nabla)\vec{v}$ due to the 
Evershed flow, while the corresponding term is negative for $B_z$ due to the 
upward decreasing vertical velocity near $\tau=1$ (see Sect. 
\ref{sect:simple-model} for more detail). \citet{Ichimoto:etal:2007} 
found also a negative $v_R - \vert B\vert$ correlation in the inner penumbra
with a trend of overall decreasing anti-correlation further out, however, a sign
change was not observed. The latter was proposed by 
\citet{Tritschler:etal:2007} and \citet{Ichimoto:etal:2008} based on 
observations of the net circular polarization (NCP) in the outer penumbra 
at different viewing angles. 

The outflow velocity we find in the simulation is not stationary, we see 
flow variability that ranges from periodic fluctuations on timescales of
$5-10$ minutes in the inner penumbra to quasi-periodic variations over a wider 
range of timescales starting from $10-20$ minutes in the center and 
outer penumbra. A flow variability in the $15-40$ minute range was also 
reported in the simulation of \citet{Kitiashvili:etal:2009} and associated
with Evershed clouds \citep{Shine:etal:1994,Rimmele:1994,Cabrera:etal:2007}. 
It is conceivable that the periodic variations 
we find near the inner tip of filaments have a relation 
to twisting motions observed by \citet{Ichimoto:etal:2007:sc} and
\citet{Bharti:etal:2010}. We focus in this paper on the maintenance of the 
stationary flow component and base our analysis primarily on time and volume 
averages over sections of the penumbra. The non stationary flow component 
will be analyzed in a separate publication.

\subsection{Mass and energy fluxes}
\begin{figure*}
  \centering 
  \resizebox{12.6cm}{!}{\includegraphics{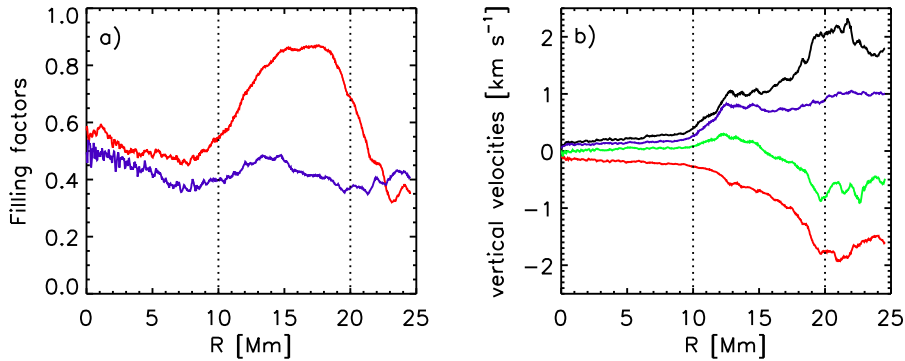}}
  \caption{a) Filling factors of radial outflows (red) and upflows 
    (blue). b) Vertical rms velocity (black), mean upflow (blue), mean 
    downflow velocity (red), and mean vertical velocity in regions with 
    outflows (green).}
  \label{fig:fill_vy}
\end{figure*}

\begin{figure*}
  \centering 
  \resizebox{12.6cm}{!}{\includegraphics{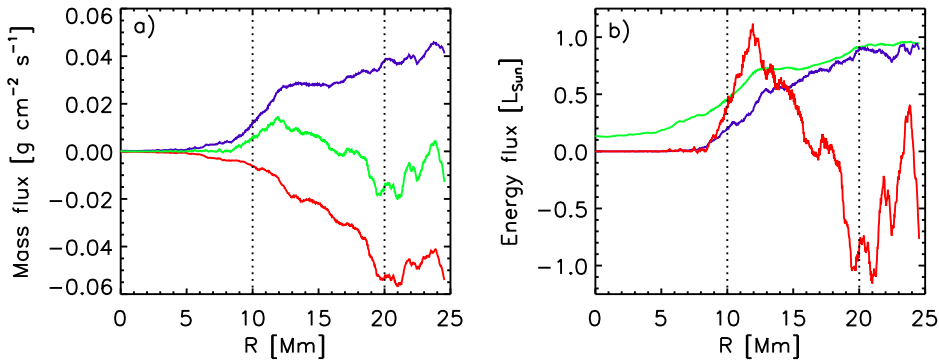}}
  \caption{a) Mass flux in upflows (blue), downflows (red), and azimuthal
    average of mass flux (green). In the inner penumbra up to $50\%$ of the 
    total upward directed mass flux is found in the large-scale flow component.
    The vertical mass flux is balanced within $R<20.9$ Mm. In this region
    about $13\%$ of the unsigned mass flux is found in the large-scale flow 
    component. b) Decomposition of convective energy flux into normalized 
    small-scale energy flux (blue) and large-scale energy flux (red). In 
    green color is shown the normalized surface intensity for comparison. 
    Integrated
    over the region with balanced vertical mass flux ($R<20.9$ Mm), the 
    contribution from the large-scale flow component is $12\%$, consistent 
    with the mass flux contribution.
  }
  \label{fig:m_e_flux}
\end{figure*}
Fig. \ref{fig:fill_vy} panel a) displays filling factors of radial
and vertical motions.
While the upflow filling factor remains almost constant around $0.4$ to 
$0.5$ from inner umbra toward the outer penumbra, the filling factor of 
outflows exceeds $0.8$ in the center of the penumbra. 
Panel b) shows the vertical rms velocity (black), together 
with the mean velocity of upflow (blue) and downflow regions (red). 
The green line presents the mean vertical velocity averaged over regions 
with radial outflows (flow channels). The latter shows a weak average 
upflow of about $250\,\mbox{m\,s}^{-1}$ in the inner penumbra
and a downflow reaching velocities of more than $500\,\mbox{m\,s}^{-1}$ toward
the outer edge of the penumbra. 

The contributions from small- and large-scale flow components to mass and 
energy flux in the penumbra are presented in Fig. \ref{fig:m_e_flux}. In 
order to properly
compare up- and downflow components we perform the analysis here on a constant
height surface that is located about $350$ km beneath $\tau=1$ in the quiet
Sun (about half a Wilson depression downward). We decompose here the
mass flux into positive and negative as well as azimuthal average components. 
Their contributions as function of radius are presented in panel a), 
where we show $\langle m_z^+ \rangle$  in blue and
$\langle m_z^- \rangle$ in red as well as $\langle m_z \rangle$ in green 
(the latter is the sum of the former two). Here, $\langle\ldots\rangle$ denotes
the azimuthal average and $m_z^{\pm}=(m_z\pm\vert m_z\vert)/2$. While in 
the innermost penumbra up to about $50\%$ of the mass flux is present in the 
azimuthal average component, this fraction drops steadily toward the center 
penumbra. Integrated over the region $R<15$ Mm about $1/3$ of the total upward 
flowing mass is found in the azimuthal component, while the major fraction 
($2/3$) is still overturning laterally. The mass flux in the penumbra is
balanced within $R<20.9$ Mm. Integrated over this region the unsigned mass flux 
in the mean component constitutes about $13\%$ of the total unsigned vertical 
mass flux in the penumbra.

Evaluating the relative contributions from large-scale and small-scale 
convective motions to the total 
convective energy flux in the penumbra requires an appropriate decomposition 
of the vertical mass flux. A separation just into azimuthal mean and the 
respective fluctuation would not be sufficient since the latter
assumes that the large-scale flow is axisymmetric and equally considers
filaments with higher temperature and the region in between with lower
temperature in the enthalpy flux. The consequence would be an underestimation of
the overall contribution from the large-scale flow (in the region $R<20.9$ 
the net contribution would be $-0.07 L_{\odot}$). 
Instead we construct the vertical mass flux of
the laterally overturning flow component $m_z^S$ as follows: in regions with
positive $\langle m_z \rangle$, we reduce the amplitude of upflows such
that they are in a mass flux balance with downflows, in regions with
negative $\langle m_z \rangle$, we reduce the amplitude of downflows such
that they are in balance with the upward directed mass flux. The large-scale
mass flux is then given by $m_z^L=m_z-m_z^S$. Unlike the decomposition into
azimuthal mean and corresponding fluctuation, this procedure does not change
the position of upflow and downflow regions for $m_z^S$ and $m_z^L$ compared
to $m_z$. With $H=(e_{\rm int}+p)/\varrho+v^2/2$ we can now compute the 
energy flux components $F_z^{S/L}=\langle m_z^{S/L} H \rangle$, which are
displayed in Fig. \ref{fig:m_e_flux}  panel b). 
While $F_z^S$ matches the intensity profile very well in the outer penumbra,
there is a clear deficit present in the inner penumbra. The deviations
in the umbra are due to the fact that our horizontal slice is located above 
the $\tau=1$ in the umbra and therefore the convective energy flux is zero. 
The large
scale energy flux $F_z^L$ has an amplitude of about $+L_{\odot}$ in the
inner penumbra and $-L_{\odot}$ in the outer penumbra. The relevant quantity
here is the net contribution after carefully balancing the upflow in the
inner and downflow in the outer penumbra. Integrating $F_z^L$ over the
region $R<20.9$ Mm (in which the large-scale mass flux is balanced) leads to
a net contribution of $12\%$ to the total convective energy flux. The latter is
very consistent with the relative mass flux contribution of $13\%$ we found
before. In an integral sense the large-scale flow contributes
only a small fraction of the energy radiated away in a sunspot penumbra,
but locally the contribution can be larger. If we use the difference between 
$F_z^S/L_{\odot}$ (blue) and $I/I_{\odot}$ (green) in Fig. \ref{fig:m_e_flux} 
as a rough estimate for the missing energy flux we identify a contribution of
up to $50\%$ in the inner penumbra. Note that we avoided in the above
discussion the association between Evershed flow and the large-scale flow
component since there is no clear definition of what the former encompasses.
If we associate the Evershed flow only with the horizontal flow pattern that 
corresponds to upflows in the inner and downflows in the outer penumbra we would
conclude that this flow pattern plays only a minor role in the penumbral energy
transport. This definition would essentially correspond to the "sources"
and "sinks" of the Evershed flow that have been identified by
\citet{Rimmele:Marino:2006} and \citet{Ichimoto:etal:2007}. Also note that
the contribution from large-scale flows increases with depth as all intrinsic 
scales of convection increase with depth. Properly quantifying their 
contribution in deeper layers requires numerical simulations over longer 
timescales (since convective timescales increase with depth), which is 
beyond the scope of the current investigation.   
   
\section{Subsurface flow structure and underlying driving forces}
\label{sect:forcings}
\subsection{Flow structure beneath penumbra}
\begin{figure}
  \centering 
  \resizebox{\hsize}{!}{\includegraphics{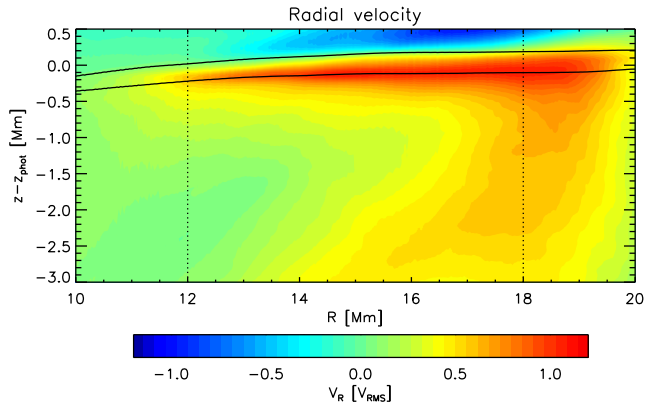}}
  \caption{Radial and depth dependence of radial flows in penumbra. The radial
    flow velocity is normalized with the rms velocity of convective motions 
    outside the sunspot (which is shown in Fig. \ref{fig:vrms} panel a). 
    The dark solid lines indicate the $\tau=1$ and $\tau=0.01$
    levels. The strongest outflows reach velocities comparable to the rms 
    velocity and are concentrated around $\tau=1$. We see a transition
    from outflow to inflow at about $\tau=0.01$ in the inner and at 
    $\tau=0.001$ in the outer penumbra. 
    In deeper layers, especially in the outer penumbra, outflows with speeds 
    reaching $50\%$ of the convective rms velocity are present. The vertical 
    dotted lines indicate 
    sub-regions we use for a more detailed analysis in the following 
    discussion.} 
  \label{fig:vr_normalized}
\end{figure}

\begin{figure*}
  \centering 
  \resizebox{12.6cm}{!}{\includegraphics{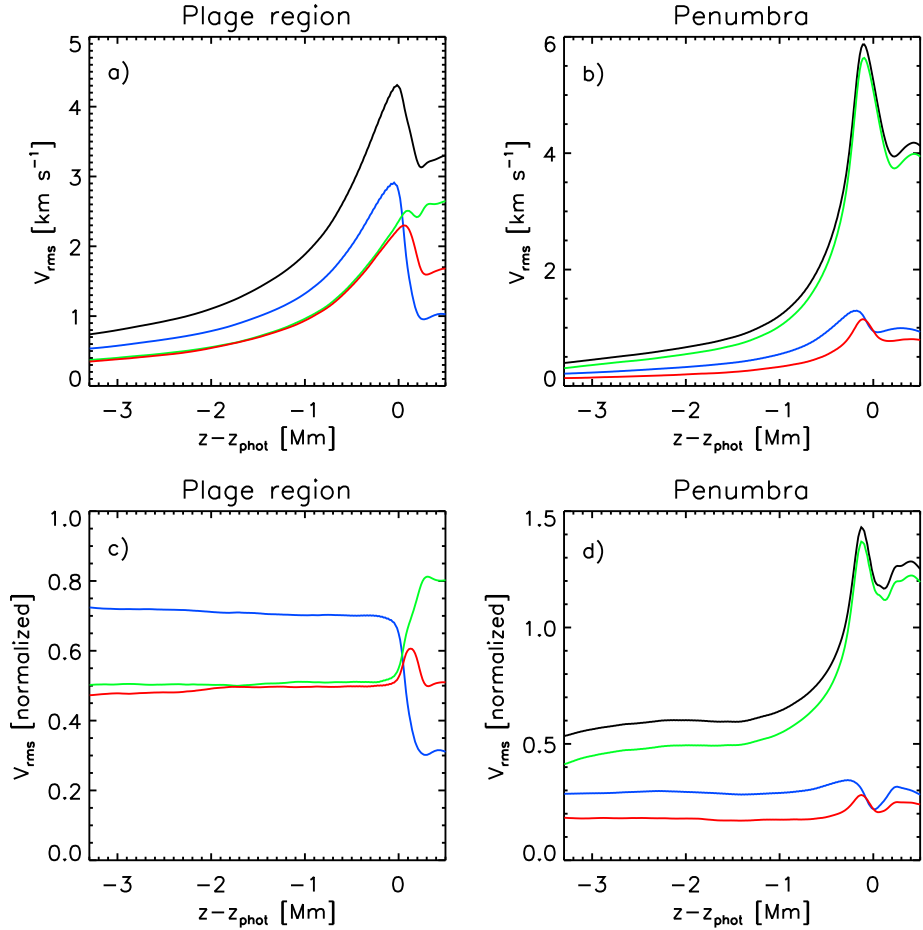}}
  \caption{Vertical profiles of rms velocities in the plage region surrounding
    the spot (a, c) and in the penumbra (b, d). The latter rms 
    velocities are computed in the center penumbra in between $R=12$ and
    $R=18$ Mm. Panels a) and b) show the absolute values, while velocities in 
    panels c) and d) are normalized by the rms 
    velocity profile outside the sunspot near the edge of the computational 
    domain. Black indicates the total, blue the vertical rms velocity, green 
    and red the 2  horizontal components. 
    In the case of the penumbra green is the component
    along the filaments. In the penumbra there are 2 distinct regimes. In the 
    deep layers ($<-1$ Mm) flows are anisotropic, but show a depth dependence
    similar to convective flows in the plage region (i.e., they scale 
    proportional to $v_{\rm rms}^0$). In the near surface layers,
    this scaling is still present for flows turning over laterally (blue and
    red curve), while
    the flow component along filaments shows a much steeper increase with 
    height.}
  \label{fig:vrms}
\end{figure*}

Fig. \ref{fig:vr_normalized} presents the subsurface outflow structure
as function of depth and radial distance from the center of the spot. The
depth is measured relative to the average height of the $\tau=1$ level
in the quiet sun. We will use the same height scale in all of the following
figures except Fig. \ref{fig:filament_avr}, where we use the average $\tau=1$
level in the penumbra as reference.
Flow velocities are normalized by the rms velocity found outside the sunspot
at the corresponding height level (see also Fig. \ref{fig:vrms}). We have
chosen this normalization in order to compare flow fields found in the penumbra 
to convective flows found in almost undisturbed convection. We refer to this 
velocity reference in the following as $v_{\rm rms}^0$. While the
outflow velocity stays around $0.4-0.5\,v_{\rm rms}^0$ in the deeper layers, the
near surface layers stand out with flow speeds exceeding $v_{\rm rms}^0$.
The two different scaling regimes of the outflow velocity found in the near 
surface layers (uppermost $500$ km) and the deeper part of the domain 
indicate already different physical driving mechanism at work,
which we will analyze further in the following discussion. We exclude here
the lower most $2$ Mm of our domain which are partially influenced by the 
bottom boundary condition. The solid black lines indicate the average 
$\tau=1$ and $\tau=0.01$ levels. The radial outflow velocity peaks close to
$\tau=1$ and falls off rapidly with height. An outflow is present to about
$\tau=0.01$ in the inner and $\tau=10^{-3}$ in the outer penumbra. The 
azimuthally averaged mass flux changes sign between $\tau=10^{-3}$ and 
$\tau=10^{-4}$, since it puts more weight on the region above the more dense
filament channels with fast outflows. Overall the simulation indicates that 
radial outflows in the 
penumbra are expected to be found in the deep photosphere, which is consistent 
with recent spectropolarimetric inversions \citep{Schlichenmaier:etal:2004,
BellotRubio:etal:2006,Borrero:etal:2008}, but not earlier work
by \citet{Rimmele:1995} and \citet{Stanchfield:etal:1997} where elevated
flow channels were inferred. Whether the inflow above $\tau=10^{-3}$ could be 
related to the inverse Evershed 
flow observed in the Chromosphere \citep{Dialetis:etal:1985} is currently an 
open question, even though also \citet{Borrero:etal:2008} found observational
evidence for an inflow near temperature minimum. As described in
\citet{Rempel:etal:2009} we switch for reasons of numerical stability to an
isothermal equation of state in regions with 
$\beta=p_{\rm gas}/p_{\rm mag}< 10^{-3}$ and limit the Lorentz force such that
the Alfv{\'e}n velocity does not exceed $60\,\mbox{km\,s}^{-1}$ to prevent 
stringent time
step constraints. The latter two could possibly influence this flow 
pattern near the top boundary while the influence on flows in the photosphere 
is rather weak. Despite substantial velocities of a few $\mbox{km\,s}^{-1}$ 
the associated mass
and momentum flux is negligible compared to photospheric flows due to the
sharp drop in density.

The vertical dotted lines
indicate 3 regions we refer to in the following analysis. We consider the
region in between $R=10$ and $R=12$ Mm as inner penumbra, the region in between 
$R=12$ and $R=18$ Mm as center penumbra and  $R=18$ and $R=20$ as
outer penumbra. We have chosen the boundary for the inner penumbra based
on the intensity profile (Fig. \ref{fig:global_prop}) that reaches a value 
typical for a penumbra of 
$0.7 I_{\odot}$ at $R=12$~Mm. Since our outer penumbra might not be fully 
representative for conditions in a "typical" outer penumbra due to the 
presence of a nearby opposite polarity spot with an Evershed flow in the 
opposite direction, we separated out regions with $R>18$~Mm. $R=18$~Mm is also
the distance where most of the dominant filaments of the center penumbra
end (see Fig. \ref{fig:int_region} and \ref{fig:fil_tau1}).

Fig. \ref{fig:vrms} compares rms velocities in the plage region surrounding
the sunspot and the center penumbra. The
top panels present the absolute rms velocities, the bottom panels relative
to $v_{\rm rms}^0$. Blue indicates the vertical rms velocity, green and
red the horizontal components (green is along the filaments in the case of
the penumbra). In the plage region (more or less undisturbed convection)
about half of the kinetic energy is found in vertical motions, the other half 
equally distributed among the horizontal components. This scaling holds very
well over the 3 orders of magnitude in pressure stratification shown here
(panel c). In the penumbra the vertical and horizontal rms velocity
perpendicular to the filaments show a similar scaling and relative strength, 
but overall their amplitude is reduced to about $40\%$ of the respective 
values in the plage region. The rms velocity along the filaments is strongly
increased with respect to the vertical rms, indicating a strong degree of
anisotropy. The rms velocity in the direction of filaments is proportional to
$v_{\rm rms}^0$ in more than $1$ Mm depth and shows a steep increase toward the 
photosphere. Overall the kinetic energy is reduced in the deeper layers, but 
doubled in the near surface layers compared to the plage region. The apparent
excess of kinetic energy found in the Evershed flow compared to the plage
region is due to a vertical redistribution of kinetic energy combined with 
anisotropy of the flow.

\subsection{Underlying driving forces}
\begin{figure*}
  \centering 
  \resizebox{12.6cm}{!}{\includegraphics{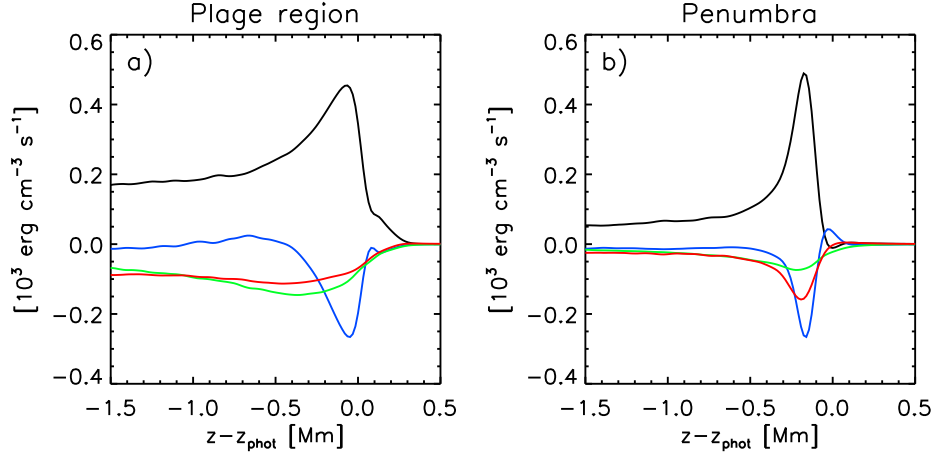}}
  \caption{Comparison of energy conversion terms in kinetic energy 
    equation between plage region (a) and center penumbra (b). 
    The quantities shown are averaged in volume and about 1 hour in time. 
    Black: work by pressure/buoyancy forces; red: work by Lorentz force; 
    blue: work by acceleration forces; green: work by viscous forces. 
    Overall the pressure/buoyancy forces are the primary driver of flows.
    Close to the surface a major fraction of the energy is deposited into
    acceleration work, viscous losses and work against Lorentz forces
    have about equal contributions.}
  \label{fig:energetics_cmp}
\end{figure*}

\begin{figure*}
  \centering 
  \resizebox{12.6cm}{!}{\includegraphics{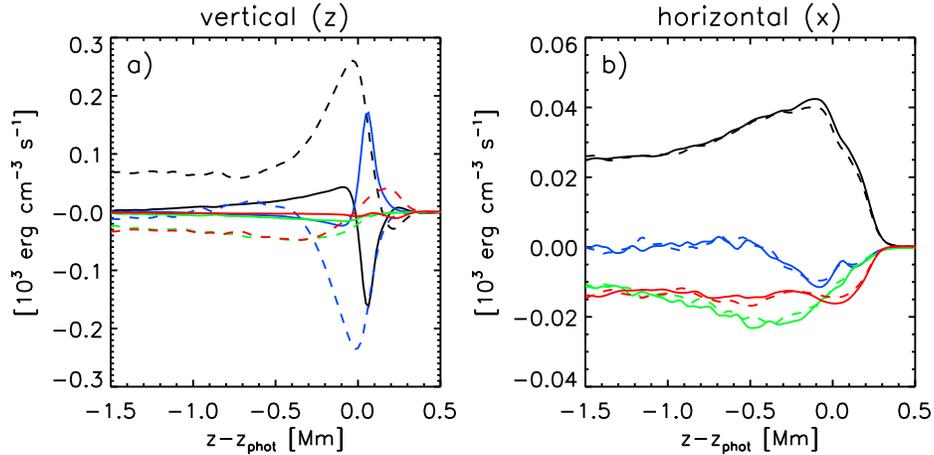}}
  \caption{Plage region: Energy conversion terms in kinetic energy 
    equation. The quantities shown are averaged in volume and about 1 hour 
    in time. Black:
    work by pressure/buoyancy forces; red: work by Lorentz force; blue:
    work by acceleration forces; green: work by viscous forces. Solid lines
    indicate upflow, dashed lines downflow regions. a) Contribution from terms
    in the vertical and b) from terms in the
    horizontal direction. In the vertical momentum equation pressure/buoyancy 
    terms are in balance with acceleration terms. Most pressure/buoyancy driving
    takes place in the downflow regions. In the horizontal momentum equations
    pressure driving is offset primarily by work against magnetic and viscous 
    forces.}
  \label{fig:energetics_plage}
\end{figure*}
\begin{figure*}
  \centering 
  \resizebox{12.6cm}{!}{\includegraphics{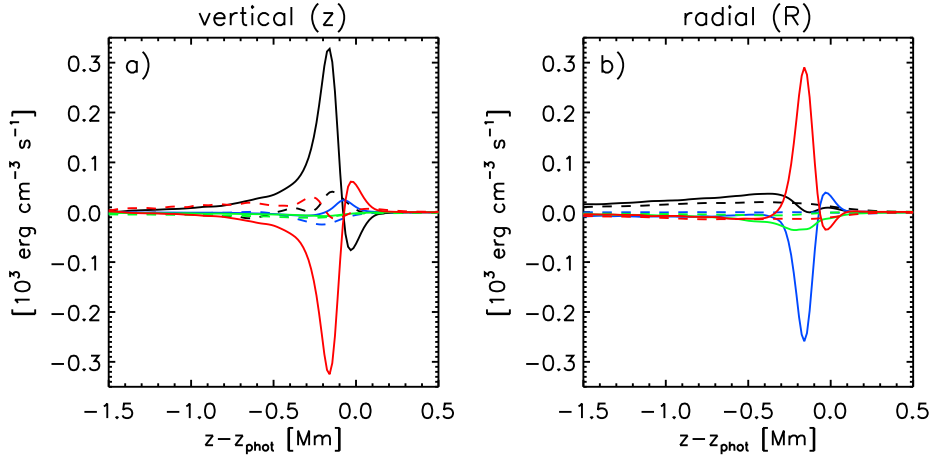}}
  \caption{Center penumbra (averaged in between $R=12$ to $R=18$ Mm): 
    Quantities shown are the same as in Fig. \ref{fig:energetics_plage}.
    In the vertical momentum equation (panel a) pressure driving is balanced
    by by work against the Lorentz force. Contrary to the plage region the 
    pressure driving takes place in upflow regions. Acceleration terms are 
    negligible. In the radial direction (panel b) outflows are primarily 
    driven by magnetic forces (in the uppermost 
    few $100$ km), which are in balance with acceleration terms. In deeper
    layers pressure driving becomes more important, but remains weaker than
    in the plage region (see also Fig. \ref{fig:energetics_center_p2}).} 
  \label{fig:energetics_center_p}
\end{figure*}

\begin{figure*}
  \centering 
  \resizebox{12.6cm}{!}{\includegraphics{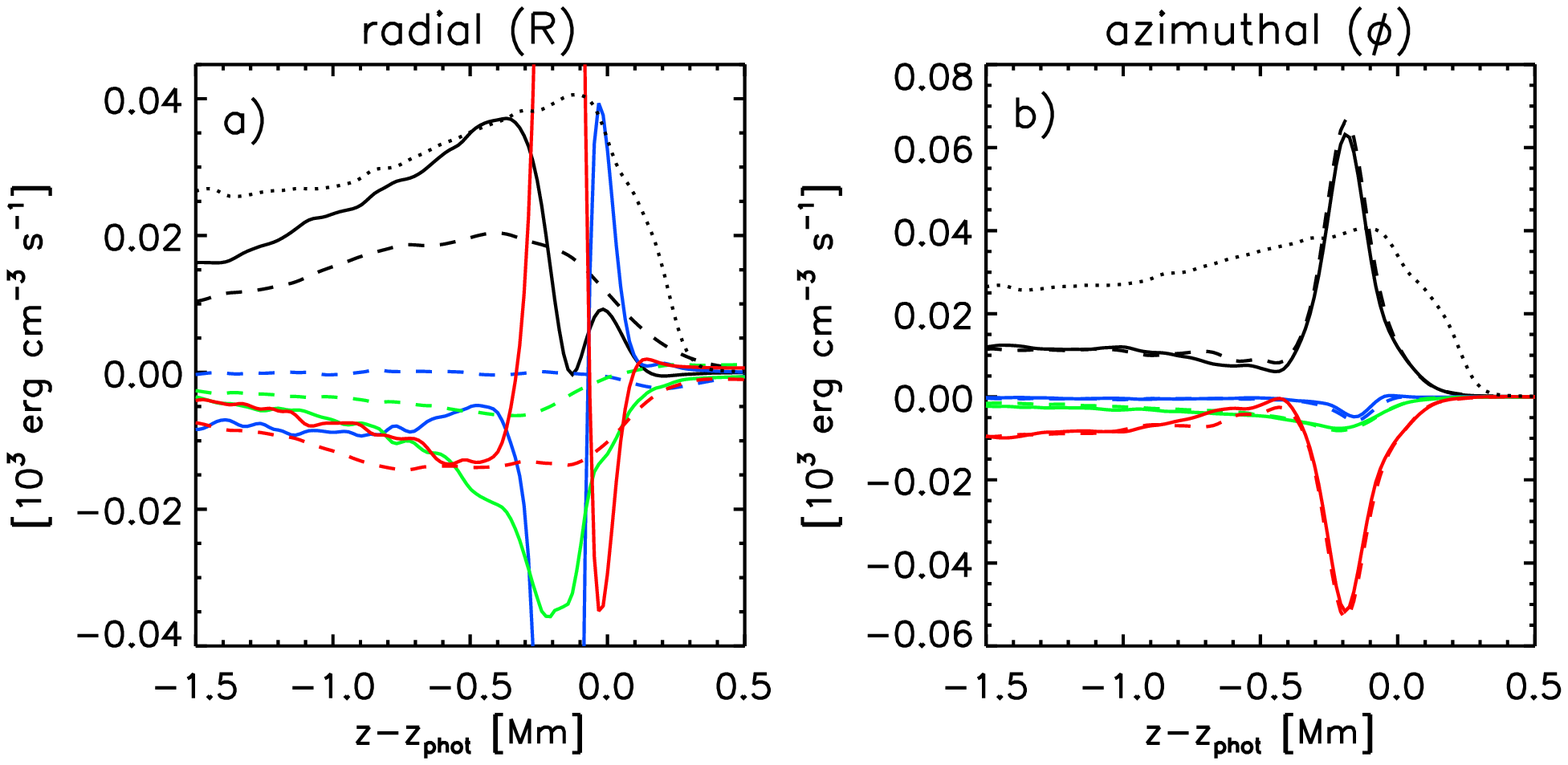}}
  \caption{Energy conversion terms for the radial direction in the center 
    penumbra on different scale (panel a). In more than $300$~km 
    depth flows are primarily driven by pressure forces. Pressure driving 
    favors outflows, the Lorentz force opposes inflows more than outflows.
    In less than $300$~km depth, the Lorentz force is the dominant driver for
    outflows, while pressure driving favors inflows. 
    Panel b) shows the balance for horizontal motions perpendicular to the 
    filaments (azimuthal direction).
    Work by pressure forces is in balance with work against magnetic forces.
    The dotted line indicates in both panels the role of 
    pressure driving in the plage region for comparison.}
  \label{fig:energetics_center_p2}
\end{figure*}

In order to investigate the physical processes that lead to the driving
of large-scale outflows around sunspots we analyze the energy conversion terms
in the kinetic energy equation. Starting from the momentum equation we derive
the following energy balance (we drop the time derivatives since we are 
interested in time averages):
\begin{equation}
\begin{split}
\underbrace{\vec{v}\cdot(\varrho\vec{g}-\nabla p)}_{\rm pressure/buoyancy}
&+\underbrace{\vec{v}\cdot(\vec{j}\times\vec{B})}_{\rm Lorentz}\\
&\underbrace{-\varrho\vec{v}\cdot[(\vec{v}\cdot\nabla)\vec{v}]}_{\rm Acceleration}
+\underbrace{\vec{v}\cdot F_{visc}}_{\rm Viscosity}=0\;.
\end{split}
\end{equation}
Under the assumption of stationarity the acceleration term is identical to
the negative divergence of the kinetic energy flux, $\varrho\vec{v} v^2/2$.
A negative acceleration term implies positive divergence, i.e. the volume
element is a source of kinetic energy.
In Fig. \ref{fig:energetics_cmp} we compare the different contributions
to the energy equation for the plage region (panel a) and penumbra (panel b).
On a qualitative level there is a large degree of similarity:
Pressure/buoyancy forces are the main driver, close to the surface most
of that energy input is used up by acceleration forces, the remainder 
is balanced in about equal parts by work against viscous and 
Lorentz forces. In the penumbra the total amount of energy input by 
pressure/buoyancy forces in the uppermost $1.5$ Mm shown here is reduced 
to about $40\%$ and more concentrated toward the photosphere.
The reduction in energy input is consistent with the overall reduced 
kinetic energy 
integrated over this depth range. It is also notable that the work against
the Lorentz force is not substantially different from the plage region
(relative to the respective pressure driving) despite the quite different
field strength and field structure.  

To investigate the driving of flows in plage and penumbra further we  
split now terms into the contributions from different grid directions
as well as flow directions, i.e. we consider the following $18$ 
terms:
\begin{eqnarray}
  P_i^{\pm}&=&\langle\,v_i^{\pm}[\varrho g_i -(\nabla p)_i]\,\rangle
  \label{eq:driving_components1}\\
  L_i^{\pm}&=&\langle\,v_i^{\pm}(\vec{j}\times\vec{B})_i\,\rangle
  \label{eq:driving_components2}\\
  A_i^{\pm}&=&-\langle\,\varrho v_i^{\pm}[(\vec{v}\cdot\nabla) \vec{v}]_i
  \,\rangle\label{eq:driving_components3}
\end{eqnarray}    
Here, $i$ indicates either the Cartesian directions $x,y,z$ in the case of the
plage region or the cylindrical components $R,\Phi,z$ in the case of the 
sunspot penumbra. With $v_i^{\pm}=(v_i\pm\vert v_i\vert)/2$ we denote
negative and positive velocity components. 
Note that we compute all forces on the Cartesian grid and use the 
transformation to cylindrical coordinates only to separate the directions
along and perpendicular to filaments in our nearly axisymmetric penumbra
fragment (i.e. we compute terms like $v_r F_r$ and $v_{\Phi} F_{\Phi}$ instead
of $v_x F_x$ and $v_y F_y$ with $\vec{F}$ being any one of the forces). 
The explicit expression for the
viscous force is rather complicated due to the non-linearity of the
underlying artificial viscosity scheme. In the following discussion we do
not explicitly compute the viscous terms, but indicate their approximate 
magnitude by using the quantity $V_i^{\pm}=-(P_i^{\pm}+L_i^{\pm}+A_i^{\pm})$.
We confirmed a close relationship for a few snapshots, for which we restarted
the code and extracted all numerical dissipation terms. 

Formally the energy conversion terms are power densities (work per volume 
and time). For the sake of making the text more readable we will refer to 
them in the following discussion very often as "work done by/against ... 
forces" instead of "work done by/against ... forces per volume and time".
Since the former is simply the latter multiplied by a unit volume element
and time interval it has no further impact on the physical meaning of these
terms.   

Fig. \ref{fig:energetics_plage} shows the energy conversion terms for
vertical motions (panel a) and horizontal motions (panel b) in the plage 
region. Note that we only 
show one horizontal direction due to isotropy. In the vertical direction 
pressure/buoyancy driving is in balance with work done against acceleration 
forces.  
Most of the pressure/buoyancy driving takes place in downflows due to their
overdense material that cannot be supported by the pressure gradient. 
Pressure/buoyancy driving in upflows is much weaker since they are very
close to a hydrostatic balance. Close to $z=0$ the sign of pressure/buoyancy 
driving is changing in upflows as a consequence of the overshoot layer
in the upper photosphere. Magnetic and viscous forces play only a minor role
in the vertical direction. Horizontal flows are primarily driven by
pressure forces. Most of the energy is absorbed by magnetic and viscous forces,
only a small amount is balanced by horizontal acceleration in the uppermost
$500$ km of the convection zone. 

\begin{figure}
  \centering 
  \resizebox{\hsize}{!}{\includegraphics{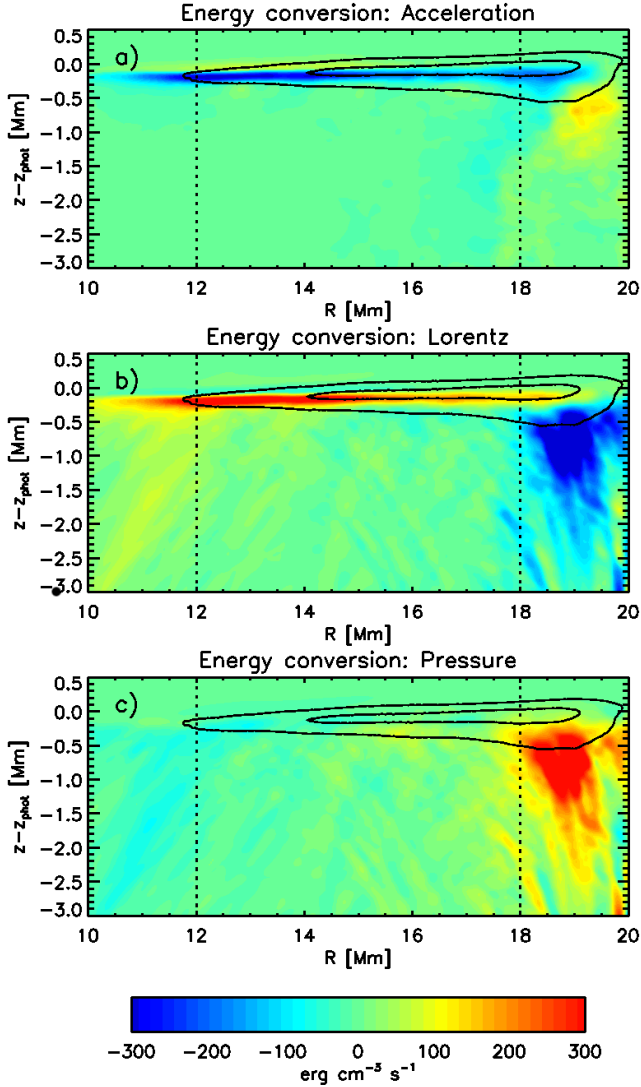}}
  \caption{Energy conversion terms as functions of radius and depth in 
    the penumbra. We show the differences between the energy conversion for
    outflow and inflow regions. Panel a) shows $A^+_R-A^-_R$, panel b)
    $L^+_R-L^-_R$, and panel c) $P^+_R-P^-_R$. The contour lines indicate 
    regions with average outflows of more than $2$ and $4\,\mbox{km\,s}^{-1}$.
  }
  \label{fig:energetics_contour}
\end{figure}

\begin{figure*}
  \centering 
  \resizebox{12.6cm}{!}{\includegraphics{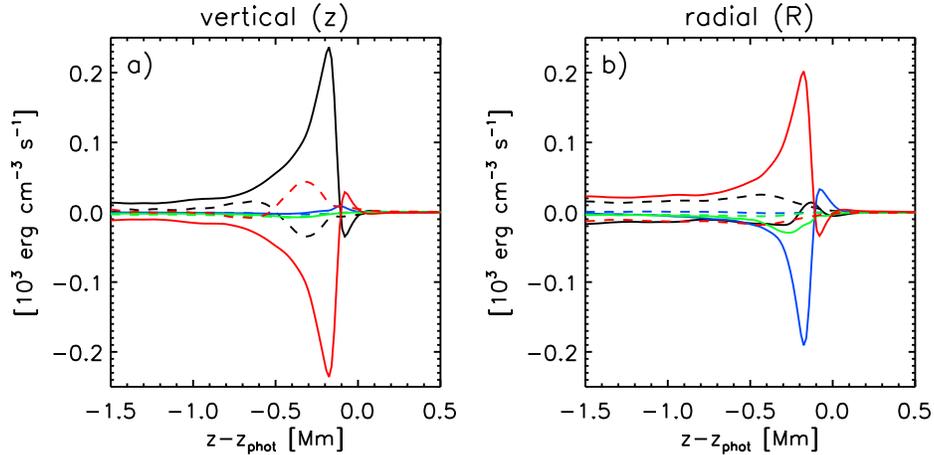}}
  \caption{Inner penumbra (averaged in between $R=10$ to $R=12$ Mm): 
    Quantities shown are the same as in Fig. \ref{fig:energetics_plage}.}
  \label{fig:energetics_inner_p}
\end{figure*}

\begin{figure*}
  \centering 
  \resizebox{12.6cm}{!}{\includegraphics{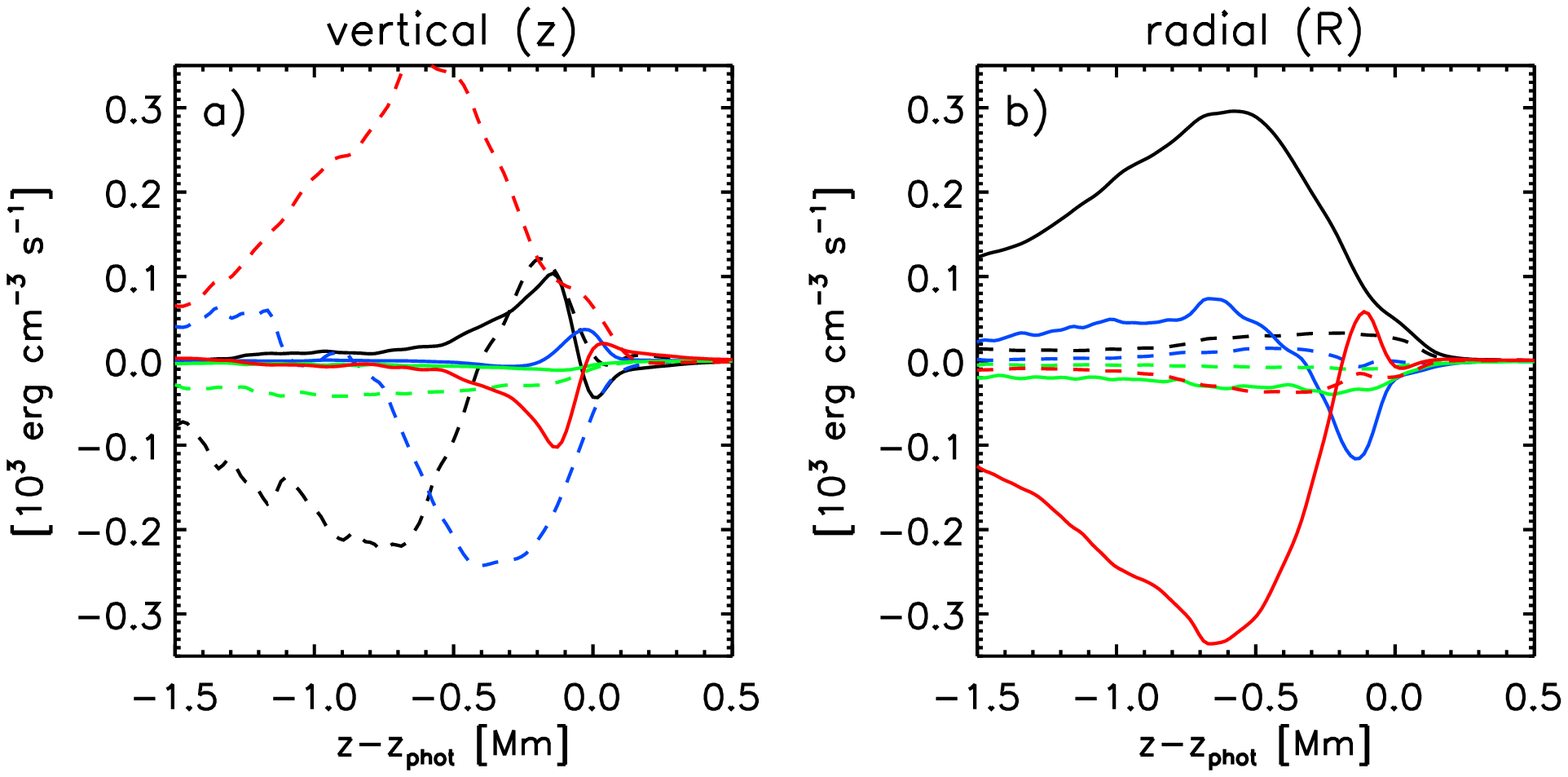}}
  \caption{Outer penumbra (averaged in between $R=18$ to $R=20$ Mm):
    Quantities shown are the same as in Fig. \ref{fig:energetics_plage}.}
  \label{fig:energetics_outer_p}
\end{figure*} 

In comparison to the plage region, the center penumbra shows distinct 
differences
(Fig. \ref{fig:energetics_center_p}). Almost all pressure/buoyancy driving
takes place in upflow regions: the presence of strong magnetic field causes 
in the near surface layers to a steepening of the pressure gradient, leading
to almost hydrostatic balance in downflows and excess pressure driving in the
upflows. This excess pressure driving in upflows is in balance with
work against the Lorentz force while in contrast to the plage region vertical 
acceleration of fluid does not play an important role at all. Most
of the energy extracted by Lorentz forces in the vertical direction is
deposited into outward acceleration of fluid along filaments. In that sense 
the Evershed flow is driven by vertical pressure forces in upflows that are
deflected into the horizontal direction through the Lorentz-force. This
"deflection" is most efficient very close to the surface: integrated from
$1$ Mm ($500$, $250$ km) depth to the top boundary about 
$44\%$ ($64\%$, $97\%$) of the pressure 
driving in the vertical direction ends up as Lorentz force driving in the 
radial direction. This way kinetic energy that is normally deposited into the 
vertical direction (plage region) gets transferred directly into the 
horizontal direction and accounts for the high anisotropy seen in the 
penumbra. 

Also the role of horizontal pressure driving differs substantially
from the plage region, which can be seen in Fig. 
\ref{fig:energetics_center_p2}a), where we show the energy conversion 
terms for the radial direction on a different scale. Pressure 
driving is dominant below $300$~km depth, but the overall magnitude 
remains smaller than in the plage region at comparable depth. 
While pressure driving  shows a preference for outflows in more than
$300$~km depth, it prefers inflows further up. Pressure forces are the primary 
cause for the deep flow component with velocities of about $0.5\,v_{\rm rms}^0$
we identified in Fig. \ref{fig:vr_normalized}, but their overall role for the 
near surface flow is limited: integrated from $1$ Mm ($500$, $250$ km) depth 
to the top boundary 
the contribution from $P^+_R$ relative to $L^+_R$ is $100\%$ ($38\%$, 
$9\%$). If we consider only the components of the driving that break the
symmetry between in- and outflows, $P^+_R-P^-_R$ and $L^+_R-L^-_R$, the 
corresponding values are $13\%$ ($1\%$, $-9\%$). Note that most of the
contribution to $P^+_R-P^-_R$ comes from the region $R>17$ Mm, further inward
$P^+_R-P^-_R$ is close to zero (see also Fig. \ref{fig:energetics_contour}). 

However, pressure forces remain the dominant driver for flows perpendicular 
($\Phi$-direction) to filaments (Fig. \ref{fig:energetics_center_p2}b). 
Here, pressure driving is 
offset by work against the Lorentz force, while both acceleration and
viscous terms do not contribute substantially. 

The different role of pressure driving compared to the plage region is 
primarily a consequence of anisotropy in terms of radially elongated 
convection cells in the penumbra: radial pressure gradients are reduced, 
while lateral pressure gradients are enhanced compared to isotropic 
granulation. In addition the steepening of the vertical pressure gradient
(that leads to the shift of pressure driving from down to upflows) results
in an overall reduction of pressure close to the photosphere in
comparison to the ambient stratification. 

The clear association between Lorentz force driving and the near surface flow
pattern is evident from Fig. \ref{fig:energetics_contour}. Here, we present
the quantities $A^+_R-A^-_R$, $L^+_R-L^-_R$ and $P^+_R-P^-_R$ as function of 
radius and depth. Strong negative values of $A^+_R-A^-_R$ indicate outward 
acceleration of fluid. These regions are confined to a narrow layer near 
$\tau=1$. Here, the Lorentz force is the primary driver, pressure
terms have weakly negative contributions (they favor inflows). The peak of
Lorentz force driving and acceleration is found in between $R=12$ and $R=15$~Mm.

In deeper layers pressure terms are in approximate balance with Lorentz force 
terms, resulting in only minor acceleration work despite their overall 
amplitude. Toward the outermost edge the Lorentz force is overcompensating 
outward directed pressure driving resulting in deceleration of radial outflows
(positive values of $A^+_R-A^-_R$).

For comparison with Fig. \ref{fig:energetics_center_p} we present in Figs.
\ref{fig:energetics_inner_p} and \ref{fig:energetics_outer_p} the same 
quantities for the inner penumbra (from $R=10$ to $R=12$ Mm) and outer 
penumbra (from $R=18$ to $R=20$ Mm). 

In the inner penumbra we see a forcing pattern that is very similar to 
that we found for the center penumbra, in particular with respect to the
near surface layers where the Evershed flow is driven. Differences are present
in deeper layers; here, the Lorentz force is also dominant in driving outflows
and is actually driving these outflows against horizontal pressure forces. 

In the outer penumbra, we see a fundamentally different situation (which is in
part a consequence of the nearby opposite polarity spot and the resulting
strongly magnetized downflow lane in between). Here, outward directed pressure 
forces dominate the picture entirely, however, they do not lead to a strong 
outward acceleration of fluid. To a large degree they are opposed by the 
horizontal Lorentz force and the energy is transferred to the vertical 
direction, where the Lorentz force becomes the major driver for downflows. 
The latter is due to the fact that magnetic field in the outer penumbra 
turns back downward. We see in the uppermost layers only a weak signal from
the horizontal Lorentz force driving outflows -- this is expected since we
are in the region where the Evershed flow declines quickly to zero.

We will further discuss the deeper reaching flow component in Sect. 
\ref{sect:deep-flow}.

\section{Magnetic filament substructure responsible for 
driving  the Evershed flow}
\label{sect:simple-model}

\begin{figure*}
  \centering 
  \resizebox{12.6cm}{!}{\includegraphics{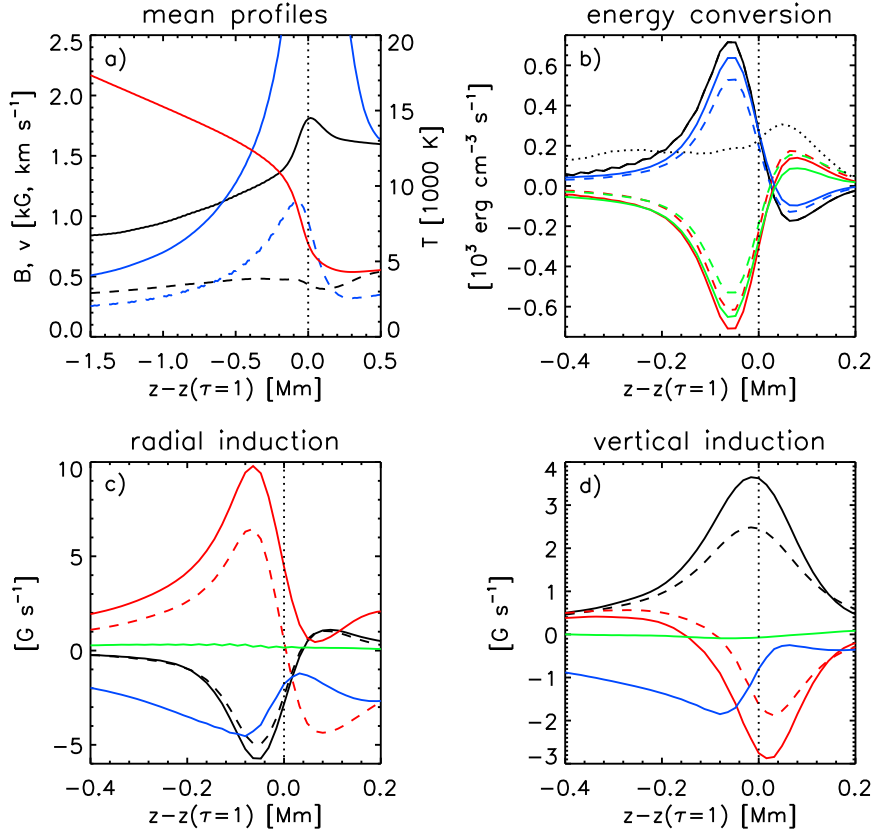}}
  \caption{a) Average vertical thermal, magnetic and flow
    profiles in regions responsible for driving the Evershed flow.
    Magnetic field (black) and flow profiles (blue) are shown on the left
    scale, solid lines indicate the radial, dashed lines the vertical 
    component. The temperature (red) is shown on the right scale. b)
    Pressure driving in upflows (black), work against Lorentz force in
    upflows (red), work by Lorentz force in radial direction (blue) and work
    against acceleration forces in radial direction (green). The dashed lines 
    show simplified expressions that are explained in the text. 
    The dotted 
    line indicates the contribution from horizontal pressure gradients
    multiplied by a factor of $10$.
    Bottom panels: Contributions in induction equation from advection (black),
    field line stretching (red), flow divergence (blue) and numerical
    diffusivity (green). Panel c) displays terms for the radial and panel d) 
    for the
    vertical field component. Here, dashed and dotted lines refer again to 
    simplified expressions explained in the text. Vertical dotted lines 
    indicate the average $\tau=1$ level.}
  \label{fig:filament_avr}
\end{figure*}

\subsection{Simplified momentum balance}
After describing the driving forces behind the Evershed flow in detail in the
previous section, we present here a simple model of the underlying
thermal, magnetic and velocity structure and reduce the overall picture to
the most relevant terms in the equations. In order to carve out the typical
structure of the regions responsible for driving the Evershed flow we select
regions which have both upflows and outflows. In Fig. \ref{fig:filament_avr}a) 
we present the mean magnetic field, flow,
and thermal structure as function of depth obtained by averaging over all 
such regions horizontally between $R=12$ and $R=18$ Mm. While the vertical 
magnetic field (black, dashed) is constant at about $400$ G, the radial 
component (black, solid) increases monotonically from about $900$ G to 
$1.8$ kG at the
$\tau=1$ level (vertical dotted line) and drops again in higher layers to
about $1.6$ kG. The steep increase just below $\tau=1$ is essential for the
Lorentz force component driving the outflow as we will describe below.
The vertical velocity (blue, dashed) increases monotonically to about $1\,\mbox{km\,s}^{-1}$
just $100$ km beneath $\tau=1$, followed by a sharp decline to a few $100\,\mbox{m\,s}^{-1}$
above $\tau=1$. The radial flow velocity peaks right at $\tau=1$,
the maximum amplitude is about $5.5\,\mbox{km\,s}^{-1}$. The solid red
line shows the mean temperature profile with the corresponding scale on the 
right. Panel b) presents volume averages of pressure driving in 
upflows (black)
work against Lorentz force in upflows (red) and work by Lorentz force
in outflows (blue) and work against acceleration forces (green) for the same 
region. The dotted black line shows contributions from horizontal 
pressure gradients multiplied by a factor of $10$. Averaged over the region 
shown (from -$0.4$ to $0.2$~Mm) they contribute about $10\%$ to the total
acceleration work. Work by the horizontal Lorentz force (blue) is $90\%$ of 
the work by vertical pressure driving (black).
The solid lines are based on all terms in the equations (see Eqs. 
(\ref{eq:driving_components1}) to (\ref{eq:driving_components3}) ), 
the dashed lines are an approximation for Lorentz force and acceleration 
terms based only on volume averages of the following expressions:
\begin{eqnarray}
  L_z&=&-v_z\frac{1}{4\pi}B_R\frac{\partial B_R}{\partial z}\label{eq:wlz} \\
  L_R&=&v_R\frac{1}{4\pi}B_z\frac{\partial B_R}{\partial z}\label{eq:wlr} \\
  A_R&=&-\varrho v_R v_z \frac{\partial v_R}{\partial z}\label{wq:war} \;.
\end{eqnarray}
The excellent agreement allows us to understand the driving mechanism behind
the Evershed effect by considering a reduced set of equations.
Note that we could go even one step further with these simplifications, by
expressing all terms through the mean quantities $\bar{B}_R(z)$, 
$\bar{B}_z(z)$, 
$\bar{v}_R(z)$, $\bar{v}_z(z)$, $\bar{p}(z)$, and $\bar{\varrho}(z)$. Despite
the fact that we are dealing with nonlinear terms of spatially highly 
inhomogeneous quantities, the agreement remains excellent except for the
acceleration term that falls short by a factor of 2, i.e. quantities
such as $\overline{\varrho v_R v_z \partial_z v_R}$ and 
$\bar{\varrho} \bar{v}_R \bar{v}_z \partial_z \bar{v}_R$ agree in general
on a qualitative level for the region we selected to perform the averages.

In the vertical direction we have essentially a magneto-hydrostatic 
balance involving the terms:
\begin{equation}
  \overline{\frac{\partial}{\partial z}\left(p+\frac{B_R^2}{8\pi}\right)}
  \approx-\overline{\varrho g}\;.
\end{equation} 
This is evident from the opposing contributions of the terms 
$-\overline{v_z\left(\partial_z p+\varrho g\right)}$ and
$-\overline{v_z B_R\partial_z B_R/(4\pi)}$ 
in Fig. \ref{fig:filament_avr}b) (black and dashed red curve). The energy 
extracted by the Lorentz force in the vertical direction leads to a strong 
acceleration of an outflow in the radial direction. Here, we have a balance
between the Lorentz force and acceleration terms:
\begin{equation}
  \overline{\frac{1}{4\pi}B_z\frac{\partial B_R}{\partial z}}\approx
  \overline{\varrho v_z\frac{\partial v_R}{\partial z}}\;.
\end{equation}
The acceleration force results from the upward transport of plasma in a 
region with an upward increasing Evershed flow velocity.
The work by vertical and radial Lorentz force components is in approximate
balance, i.e. 
\begin{equation}
  \overline{v_z\frac{1}{4\pi}B_R\frac{\partial B_R}{\partial z}}\approx
  \overline{v_R\frac{1}{4\pi} B_z\frac{\partial B_R}{\partial z}}\;,
\end{equation}
leading to a simple relation between vertical and radial flow velocities
of the form
\begin{equation}
  \overline{v_z B_R}\approx \overline{v_R B_z}\label{eq:vrvz-ratio}\;.
\end{equation}
The latter is the relation one would expect from a simple "deflection"
of vertical flows by an inclined magnetic field.

Note that most of the acceleration of the outflow takes place about $100$ km
beneath the $\tau=1$ level, while the outflow peaks right at $\tau=1$. The 
latter is a consequence of the strong vertical upflow advecting accelerated 
fluid a few $100$ km further upward. This upward advected fluid also overpowers
the inward directed Lorentz force found right above $\tau=1$ due to the
sign change in $\partial_z B_R$. 

\begin{figure*}
  \centering 
  \resizebox{15cm}{!}{\includegraphics{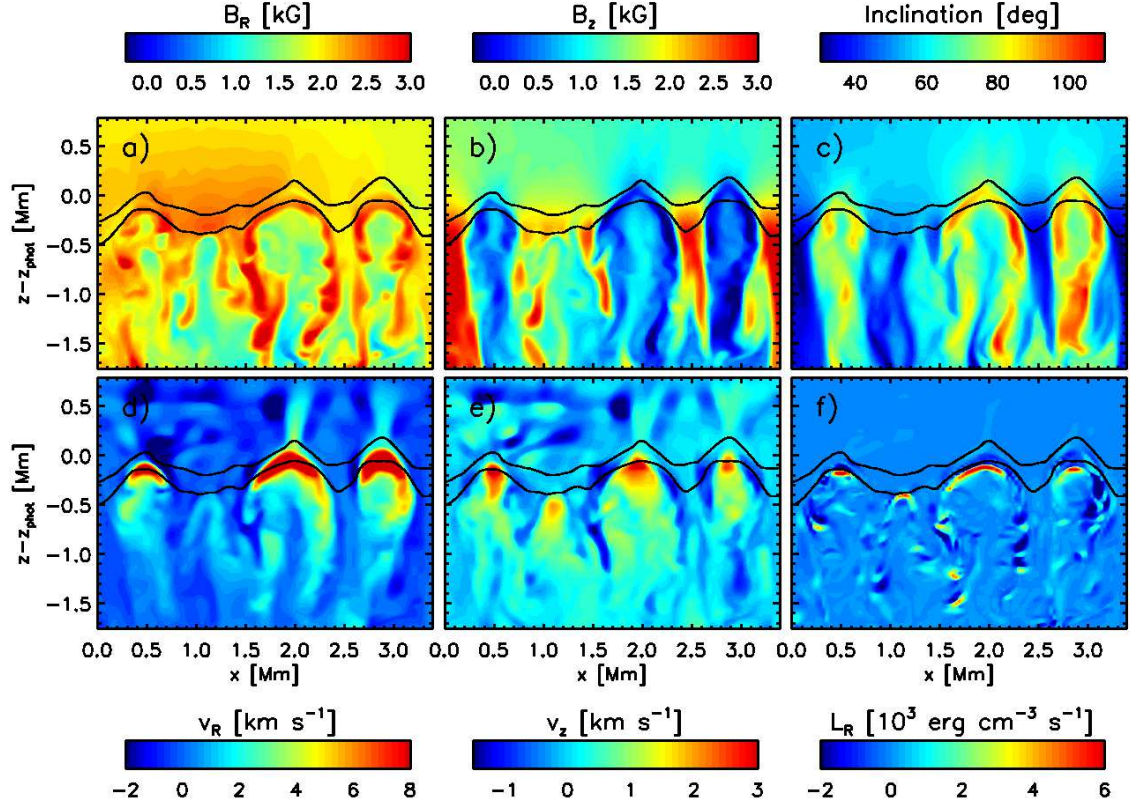}}
  \caption{Vertical cross section of filaments in the inner penumbra.
    Displayed are a) radial and b) vertical field strength together with the 
    resulting inclination in panel c). The bottom panels show d) radial, e) 
    vertical velocity and f) the energy conversion by the horizontal
    Lorentz force along filaments. The two solid lines indicate the $\tau=1$ 
    and $0.01$ levels. The enhancement of $B_R$ combined with a 
    reduction of $B_z$ around $\tau=1$ leads to a sharp increase of the 
    inclination angle. The strong increase of the inclination angle is 
    restricted to a very narrow boundary layer near $\tau=1$, around which
    we also see the dominant contribution from the horizontal Lorentz force.
    The resulting outflow found around $\tau=1$ is broader than the boundary 
    layer, but also restricted to the deep photosphere.}
  \label{fig:fil_cross_sect}
\end{figure*}

\subsection{Induction equation}
Since the large positive value of $\partial_z B_R$ right below $\tau=1$ 
plays an essential role in the acceleration process, we analyze now how this
magnetic field structure is maintained in the presence of strong vertical and 
radial flows. To this end we evaluate the different contributions in the 
induction equation:
\begin{equation}
  \frac{\partial \vec{B}}{\partial t}=
  \underbrace{-(\vec{v}\cdot\nabla)\vec{B}}_{\rm Advection}
  +\underbrace{(\vec{B}\cdot\nabla)\vec{v}}_{\rm Stretching}
  \underbrace{-\vec{B}(\nabla\cdot\vec{v})}_{\rm Divergence}\;.
\end{equation}
The bottom panels of Fig. \ref{fig:filament_avr} present these contributions
(black: advection, red: stretching, blue: divergence) for the maintenance of 
the radial magnetic field structure in panel c) and vertical magnetic field 
structure in panel d). As before solid lines show the full expressions, 
dashed-line approximations are described in the text below. 
For the radial field component (panel c) the dominant source is the
stretching term in the induction equation. The major contribution 
to this term comes from the vertical shear profile of the Evershed flow, 
leading to an induction term $\overline{B_z\partial_z v_R}$
(red dashed line). The remainder is due to horizontal stretching from terms like
$\overline{B_R\partial_R v_R}$. The peak of the stretching term 
(including all contributions) is located about $100$ km beneath $\tau=1$, 
where we also find the peak of the Lorentz force driving. This is
not exactly where we find the peak of $\bar{B}_R$, since there is an additional 
strong contribution from the vertical advection that pushes strong radial 
field upward ($-\overline{v_z\partial_z B_R}$, black dashed line). The 
remainder is offset by the negative contribution from the diverging convective 
motions (solid blue line). In the case of the vertical field (panel d), the 
role of the contributions from advection
and stretching are opposite. Here ,advection is the primary mechanism that
maintains the field. The positive sign originates primarily from horizontal
advection terms (black dashed) with a dominant contribution from 
$-\overline{v_R\partial_R B_z}$ due to the on average outward decreasing 
vertical field strength, but there are also positive contributions from 
vertical advection $-\overline{v_z\partial_z B_z}$. The dominant negative 
contribution to the stretching term is due to $\overline{B_z\partial_z v_z}$ 
(red dashed line), which peaks close to $\tau=1$ where $\partial_z \bar{v}_z$ 
is strongly negative. The remainder is offset again by the negative 
contribution from the diverging convective motions (solid blue line).
In both panels the green curve
indicates the negative sum of these three terms, i.e. the amplitude of 
additional  contributions from artificial numerical diffusivity. For both 
radial and vertical magnetic field the contributions from stretching, 
advection and divergence are in balance at the level of a few $\%$. 
This indicates that the magnetic structure within the penumbral filaments 
in this simulations is not strongly affected on average
by the unavoidable artificial magnetic diffusivity of the numerical scheme.
Also the fact that our simulation contains almost field free umbral dots on
scales even smaller than filaments sets strong constraints on the role
artificial diffusivity plays.

\begin{figure*}
  \centering 
  \resizebox{12.6cm}{!}{\includegraphics{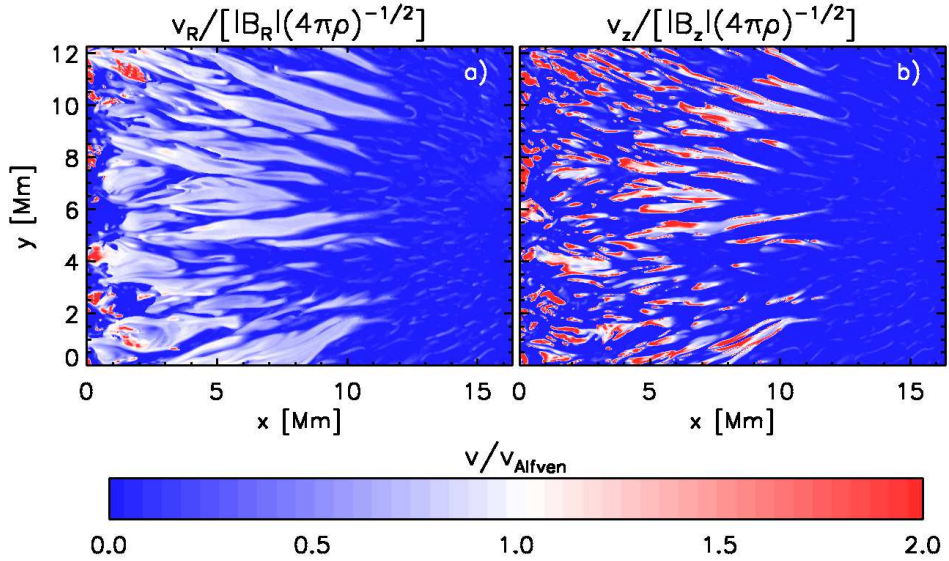}}
  \caption{Displayed are a) outflow and b) upflow velocities at $\tau=1$
      relative to $\vert B_R\vert/\sqrt{4\pi\varrho}$ and 
      $\vert B_z\vert/\sqrt{4\pi\varrho}$, 
      respectively. Blue (white, red) colors indicate sub-Alfv{\'e}nic
      (Alfv{\'e}nic, super-Alfv{\'e}nic) flow speeds. Throughout most of
      the penumbra the outflows stay close to 
      $0.8\,\vert B_R\vert/\sqrt{4\pi\varrho}$.}
  \label{fig:v_alf}
\end{figure*}

\subsection{Filament cross section} 
From Fig. \ref{fig:filament_avr} we deduce a vertical extent of about 
$200$~km for the region in which most of the energy conversion takes place. 
This value is obtained through an average over all areas between $R=12$ and 
$R=18$~km that have upflows and outflows.
Since the typical height variation of the $\tau=1$ level in the penumbra is 
about $200$~km, this indicates that locally within individual filaments the 
flow is driven in an even narrower boundary layer right beneath the $\tau=1$ 
level. We illustrate this in Fig. \ref{fig:fil_cross_sect}, which shows
magnetic field and velocity together with inclination and energy conversion 
by horizontal Lorentz force on a vertical cut through three developed and one
just forming penumbral filament in the inner penumbra. It is evident that there 
is a narrow boundary layer forming along the $\tau=1$ surface that is 
characterized by increased $B_R$ and reduced $B_z$, resulting in a strong 
increase of inclination. Lorentz force 
driving is concentrated to an equally thin layer just beneath $\tau=1$, the 
resulting outflow is broader and extends above $\tau=1$. The latter is a 
consequence of the presence of overturning motions that transport and 
distribute accelerated fluid above and along the $\tau=1$ surface. This
redistribution does not require additional acceleration work, since
the associated kinetic energy flux is close to divergence free (the term
for acceleration work is identical to the negative divergence of the kinetic
energy flux under the assumption of stationarity). In Fig. 
\ref{fig:fil_cross_sect}, we highlighted a cross section in the inner penumbra,
further out the filamentary structure is less prominent. Nevertheless, we also
see there a concentration of the energy conversion terms to very thin sheets
beneath $\tau=1$, while fast outflows are found mostly between $\tau=1$ and
$\tau=0.01$. In a statistical (average) sense the differences between inner 
and center penumbra are small (compare Figs. \ref{fig:energetics_center_p} 
and \ref{fig:energetics_inner_p}). Also note that the concentration of energy
conversion by Lorentz forces to thin sheets is typical for magnetoconvection 
in a more general sense; however, the preferred location and quasi-steady 
maintenance of these regions near $\tau=1$ is restricted to the penumbra. The 
mechanism responsible for the latter is explained in Sect. 
\ref{sect:boundary_layer}.

\subsection{Formation of thin boundary layer}
\label{sect:boundary_layer}
In this section we illustrate the crucial role of the vertical 
advection terms by discussing a simplified model that captures the
essential terms on a qualitative level to within a factor of two. We consider
only those terms in the radial momentum and induction equations that have 
been identified as the dominant contributors in the previous discussion:    
\begin{eqnarray}
  \frac{\partial v_R}{\partial t}+v_z\frac{\partial v_R}{\partial z}&=&
  \frac{B_z}{4\pi\varrho}\frac{\partial B_R}{\partial z}
  \label{eq:eq-m}\\
  \frac{\partial B_R}{\partial t}+v_z\frac{\partial B_R}{\partial z}&=&
  B_z\frac{\partial v_R}{\partial z}\label{eq:eq-i}\;.
\end{eqnarray}
Neglecting the advection terms, Eqs. (\ref{eq:eq-m}) and (\ref{eq:eq-i}) 
lead to wave solutions of the form (assuming that $B_z$ is nearly constant,
which is at least true for the average shown in Fig. \ref{fig:filament_avr}):
\begin{equation}
  \left[\frac{\partial^2}{\partial t^2}-\frac{B_z^2}{4\pi\varrho}
    \frac{\partial^2}{\partial z^2}\right](v_R,B_R)=0\;.
\end{equation}
In this case the profiles of $v_R$ and $B_R$ could not be 
maintained in place and would spread out with the Alfv{\'e}n velocity
corresponding to the vertical magnetic field component, 
$\vert B_z\vert /\sqrt{4\pi\varrho}$; see, for example, 
\citet{Vasil:Brummell:2009} for 
a discussion of the dynamics of magnetic shear layers. 
Since the Alfv{\'e}n velocity is of the order of $2\,\mbox{km\,s}^{-1}$
(using a mean value of $B_z=400$ G and 
$\varrho=3\cdot 10^{-7} {\rm g}/{\rm cm}^3$ 
near the $\tau=1$ level), the rather narrow Evershed flow profile would 
broaden substantially within a few $100$ secs of time. On the other hand, 
the inclusion of the advection terms allows for a stationary solution 
provided that $v_z>\vert B_z\vert/\sqrt{4\pi\varrho}$.
Since near $\tau=1$ the vertical field strength drops and upflows can reach
locally up to $3\,\mbox{km\,s}^{-1}$, this condition can be met within the 
thin boundary layer in which most of the driving is taking place. 
With a sufficiently strong advection term the upflow
counteracts the downward traveling Alfv{\'e}n wave, while the upward traveling 
wave is bound by the photosphere and transition to a low $\beta$ regime. The 
quasi-steady maintenance of the shear layer despite the back-reaction of 
Lorentz 
forces seems to be at odds with Lenz's rule, however, we have to keep in mind 
that there is a steady flow of energy through the system ultimately driven by
overturning convective motions. Indeed, the energy conversion by the vertical
advection term in the induction equation, $B_R/(4\pi)\, v_z\partial_z B_R$, is 
identical to the energy extracted by the Lorentz force from convective motions 
in the vertical direction (Eq. \ref{eq:wlz}).

To summarize, the most important feature responsible for driving the Evershed
flow is a strong increase of $B_R$ just beneath the $\tau=1$ 
level (combined with the presence of a vertical background field of a few 
$100$ G). The steep gradient of $B_R$ is primarily maintained by 
the vertical shear profile of the Evershed flow in combination with upward 
advection due to the strong upflow in the center of the filament. 
The vertical advection terms in the 
induction as well as momentum equation are essential for a quasi-stationary 
configuration with lifetimes far beyond the overturning timescale of 
convection. They ensure that the peak of $v_R$ and $B_R$ is
maintained above the region with the strongest Lorentz force driving as well
as induction due to shear, which are proportional to $\partial_z B_R$ and
$\partial_z v_R$, respectively. The vertical confinement of the shear
layer is guaranteed by the fact that locally the upflow velocity can
exceed $\vert B_z\vert/\sqrt{4\pi\varrho}$. Since the outflow velocity is linked
to the upflow velocity by the approximate relation $v_R B_z\approx v_z B_R$
the expectation is that the resulting outflow reaches a velocity of about
$v_R=\vert B_R\vert/\sqrt{4\pi\varrho}$. With $B_R=1\ldots 2$~kG and 
$\varrho=3\cdot 10^{-7}{\rm g}{\rm cm}^{-3}$ the resulting velocities should be
$v_R=5\ldots 10\,\mbox{km\,s}^{-1}$, which is about the range we find for the  
velocity within flow channels. Fig. \ref{fig:v_alf} displays outflow and
upflow velocities at $\tau=1$ relative to $\vert B_R\vert/\sqrt{4\pi\varrho}$ 
and $\vert B_z\vert/\sqrt{4\pi\varrho}$. Outflows are close to Alfv{\'e}nic 
throughout the penumbra (typically 
$v_R\approx 0.8\,\vert B_R\vert/\sqrt{4\pi\varrho}$), 
upflows are weakly super-Alfv{\'e}nic.

The fact that there is a clear threshold for the onset of this driving
mechanism ($v_z>\vert B_z\vert/\sqrt{4\pi\varrho}$ near $\tau=1$) is a possible 
explanation for a more or less well defined inner edge of the penumbra,
or related, a critical field inclination that needs to be exceeded (about $45$
degrees in this simulation). Going further inward toward the umbra the
vertical field becomes stronger while vertical velocities are reduced,
which makes it harder to pass this threshold locally. Even if it would be
passed the resulting radial flow velocities would be less due to the
smaller value of $B_R$.

\subsection{Observable consequences}
Unfortunately the "feature" responsible for driving the Evershed flow remains
well hidden beneath the $\tau=1$ level. Even worse, if we compute the Lorentz 
force from the "visible" part of the magnetic field structure, the Lorentz 
force is inward directed due to the sign change of $\partial_z B_R$. 
The amplitude of the visible inward component above $\tau=1$ is about a 
factor of $5-10$ smaller than the strong outward directed component beneath 
$\tau=1$, which is responsible for the outward acceleration.

The observable parts of the magnetic and velocity field are
an increase of $v_R$  and $B_R$ toward 
$\tau=1$ and a strong vertical gradient of $v_z$ around $\tau=1$. 
The latter is only visible very deep in the photosphere. The combination of
this three factors should lead to positive values of 
${\rm d}|B|/{\rm d}\tau$ and ${\rm d}|v_{\rm los}|/{\rm d}\tau$ for a 
variety of observation angles and therefore contribute to the net circular 
polarization observed in sunspot penumbrae.  

\begin{figure*}
  \centering 
  \resizebox{\hsize}{!}{\includegraphics{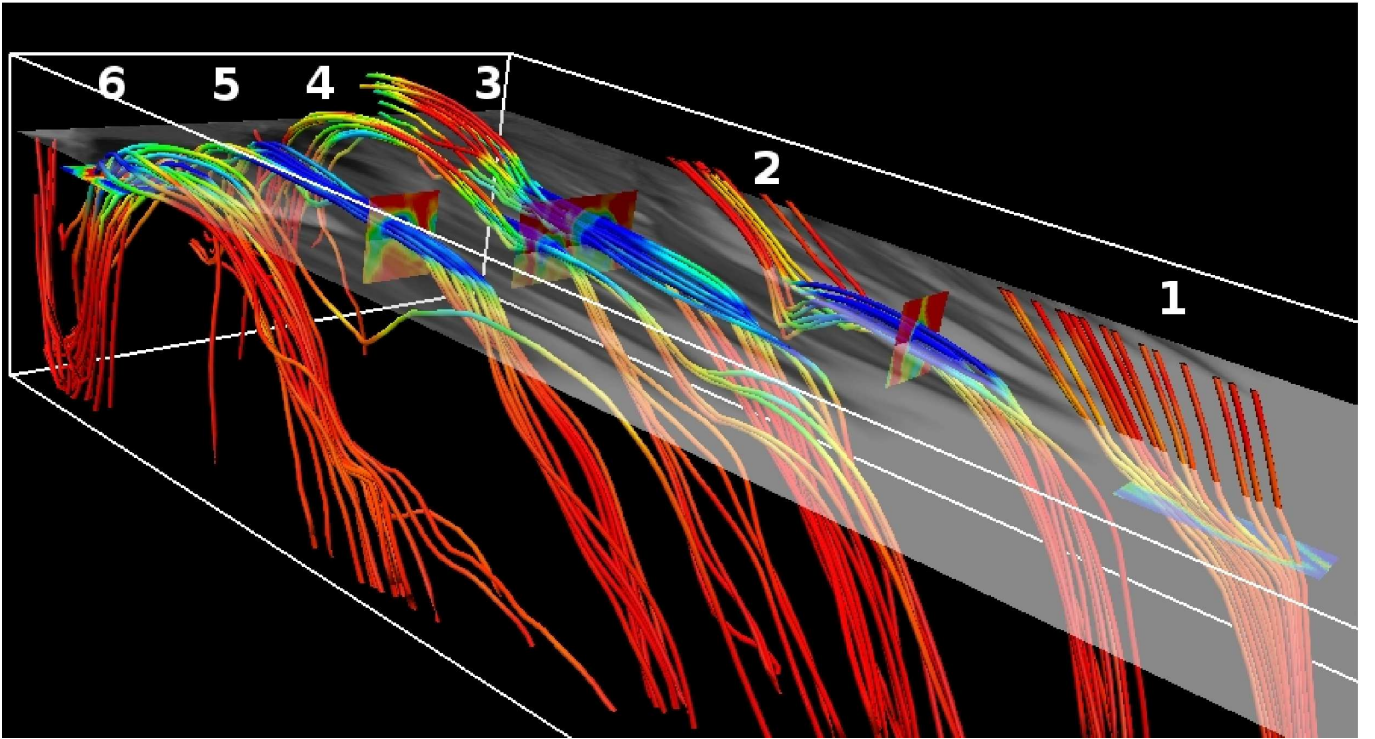}}
  \caption{Field line connectivity and associated horizontal flow speeds
    in simulated penumbra. The color of the field lines indicates the radial
    flow velocity (the colors red, yellow, green, and blue correspond to 
    velocities 
    of $<0$, $2$, $4$, and $>8\,\mbox{km\,s}^{-1}$, respectively). Filament 1 
    indicates a peripheral umbral dot almost transitioning to a penumbral 
    filament. Filaments 2-6 sample different radial position in the penumbra.
    The semi-transparent
    plane indicates a magnetogram near (average) $\tau=1$. Smaller horizontal
    and vertical cross section indicate the regions from which we selected 
    the seed points for the field line integration.
  }
  \label{fig:field_lines}
\end{figure*}

\section{Field line structure of filaments}
\label{sect:filaments}
Many simplified models of penumbral filaments and the origin of the Evershed 
flow such as 
\citet{Meyer:Schmidt:1968,Thomas:1988,Degenhardt:1989,Degenhardt:1991,
Thomas:Montesinos:1993,Montesinos:Thomas:1997,Schlichenmaier:etal:1998a,
Schlichenmaier:etal:1998b} 
are based on the thin flux tube approximation. It is not clear from first 
principles whether penumbral filaments (flow channels) can be identified with
flux tubes in a meaningful way. The latter assumes the existence of a well 
defined flux surface that encloses a flow channel and clearly separates 
fluid "inside" the channel from fluid "outside" and assumes further that there 
are well defined "footpoints" when intersected with a horizontal plane 
somewhere beneath the photosphere.

The convective structure of the penumbra as presented in Sect. 
\ref{sect:phot} puts already some limits on the usefulness of the flux tube
concept, since overturning convective motions are mostly orthogonal
to the flows assumed in the flux tube picture (except footpoints).
These convective motions lead to a continuous mass, momentum and energy
exchange while fluid is moving outward along penumbral flow channels.
Integrated over the penumbra this mass exchange is substantial, since we find
only about $13\%$ of the total unsigned vertical mass flux in the large
scale flow component.

In the following paragraphs we will discuss filaments on the basis of their
field line structure and connectivity to allow for a better comparison
with models that are based on the flux tube approximation.

\subsection{Field Line connectivity}
In Fig. \ref{fig:field_lines} we present the magnetic field line 
connectivity as well as radial outflow velocity. The filaments are 
representative for different radial positions in the simulated penumbra. 
The field line analysis
presented here was performed using the VAPOR software package developed at NCAR
\citep{Clyne:Rast:2005,Clyne:etal:2007} (www.vapor.ucar.edu). 
The field lines are computed from a $15$ minute average, which is about a 
characteristic wave crossing time along filaments. We have chosen the latter 
since in particular stationary flux tube models make only sense on timescales
beyond that, but the following conclusions are not affected by the 
averaging in a fundamental way. Filament 1 corresponds to a peripheral
umbral dot that almost transitions to a penumbral filament, filament 2 is
representative for the inner, 3 and 4 for the center and 5 and 6 for the
outer penumbra. Seed points for filament 1 were chosen from a magnetogram at
the umbral $\tau=1$ level (regions with reduced field strength indicating
peripheral umbral dot). Seed points for filaments 2-5 were chosen based on 
outflow velocities in the indicated vertical cross sections with more than 
$5\,\mbox{km\,s}^{-1}$ (most of them are actually around 
$8\,\mbox{km\,s}^{-1}$). The seed points for 
filament 6 were chosen from the indicated horizontal plane based on regions 
with more than $5\,\mbox{km\,s}^{-1}$ downflow speed. The color coding of the 
field lines 
indicates the radial outflow velocity (the colors red, yellow, green, and blue 
correspond to velocities of $<0$, $2$, $4$, and $>8\,\mbox{km\,s}^{-1}$, 
respectively). 
Going radially outward from filament 1 to filament 6 we see the following 
physical picture emerging. Near the umbra-penumbra boundary upflow plumes
(similar to those forming umbral dots in the center of the umbra) push mass
upward along inclined field lines. The mass loading results in a small bend
of the field line and the upflow is guided outward leading to outflow
velocities of about a $\mbox{km\,s}^{-1}$. Due to the strong almost vertical 
field the flow
is not powerful enough to bend over field lines completely. This is consistent
with a moderate enhancement of the field inclination in peripheral umbral 
dots by some $10-20\deg$ that has been inferred from spectropolarimetric 
observations
\citep{Socas-Navarro:etal:2004,Riethmueller:etal:2008,Sobotka:Jurcak:2009,
Ortiz:etal:2010}. Going further outward
(filament 2) field lines are bent over sufficiently to become horizontal
for a distance of a few Mm. In the horizontal stretch we find outflows 
exceeding $8\,\mbox{km\,s}^{-1}$ (blue color), nevertheless, the field lines 
remain
connected to the top boundary. Near the outer edge of this filament, we see
the formation of small dips right before the field lines connect back to the
top boundary. The latter is a consequence of the mass flux decoupling on
average from the field through the formation of U-loops and reconnection with
deeper field lines that extend further out. In addition, downflows present
along the edge of filaments can transport field lines together with the mass
flowing along them downward beneath the photosphere.  
Moving to the center penumbra (filaments 3 and 4) the horizontal stretch with 
fast outflows is extended and field lines start to bend over and return 
beneath the photosphere (filament 4). Here, most of the mass unloading from 
field lines happens through either the above mentioned U-loop formation or 
through field lines that bend over temporarily. Going further
outward (filament 5) the latter scenario happens most of the time resulting
in a filament that follows the $\tau=1$ level and returns back beneath the 
photosphere. Note that all of these filaments (2-5) have fast $>8$~km outflows
along their almost horizontal stretch in the photosphere regardless of their
connectivity further out.
Filament 6 shows an example of field connectivity and flows in proximity of 
one of the fast downflow patches in the outer penumbra
(all of the selected field lines have more than $5\,\mbox{km\,s}^{-1}$ 
downflow velocity
near $\tau=1$). Compared to the previous filament most field lines host
only outflows in the $2-4$ km range with some faster flows present in the
ultimate proximity of the footpoint. Most field lines reach toward the higher 
photosphere and turn back beneath the photosphere within $3$ Mm. They do not 
show the extended horizontal stretches that host fast outflows in 
filaments 2-5.

\subsection{Underlying physical picture}
Overall we do not see compelling evidence that the field line connectivity
is linked to the presence of strong horizontal outflows. Filaments with
more than $8\,\mbox{km\,s}^{-1}$ outflow speed can have any connectivity: 
in the inner
penumbra field lines typically connect to the top boundary, in the outer 
penumbra field lines bend over and return beneath the photosphere. Looking
at the smooth transition in the filament structure throughout the penumbra 
(filaments 2-5) strongly suggests that a similar process 
is responsible for driving outflows in all of them. Following up on the
discussion in Sect. \ref{sect:forcings} and \ref{sect:simple-model} outflows
result from pressure driven upflows that load field lines with mass and bend
them over. A very similar situation was already described by 
\citet{Scharmer:etal:2008} based on the "slab"  simulation of 
\citet{Heinemann:etal:2007}.
The energetic signature of this process is a balance between 
pressure and Lorentz forces in upflows and Lorentz and acceleration forces
in the radial direction. Horizontal pressure gradients do not enter the picture
on average since the upflow cells are elongated in the radial direction, which 
strongly decreases the role of the radial pressure gradient. The elongation
of filaments does not impact the Lorentz force since there only vertical 
field gradients matter (see Sect. \ref{sect:simple-model}). The region in
which outflows are driven is confined to a narrow boundary layer just 
beneath $\tau=1$ as we discussed in Sect. \ref{sect:simple-model}. 
An interesting property of this driving mechanism is that no substantial 
acceleration work is present in horizontal stretches of the magnetic field,
where we find most of the fast horizontal flows: the Lorentz force
has no horizontal component there and horizontal pressure gradients do not 
contribute much. This is however no contradiction, since all of the fluid that
appears at or above $\tau=1$ has to pass through the narrow boundary layer
with concentrated driving forces (see Fig. \ref{fig:fil_cross_sect}). The 
continuous vertical mass exchange along the flow channel maintains an almost 
steady flow despite the fact that no substantial driving forces exist above 
$\tau=1$. 

The driving of outflows depends primarily on conditions in the upflow cell 
in the inner penumbra, the field line connectivity toward the outer penumbra 
is secondary and established as a consequence of the outflow (similarly 
also umbral dots form initially on field lines that might connect to a region 
several $100$~Mm away, the field line connectivity changes as part of
the process until overturning convection is possible). The average field line 
connectivity found in the penumbra depends on the radial position
(combination of ambient field strength and inclination angle). 
Mass unloading happens primarily through U-loop formation and reconnection in 
the inner penumbra since flows are not strong enough to completely bend over
field lines. In addition, entire field lines can be submerged by laterally 
overturning convective motions. In the center and outer penumbra, continuous 
bending of the field lines increases the length of the almost horizontal 
part near $\tau=1$ until mass can be unloaded at the outer edge of the 
penumbra. This process happens
periodically in the center and permanently over the life time of the filament 
in the outer penumbra. The fact that the flow speed in the flow channels does
not show an increase when field lines bend over completely is a strong 
indication that the conditions of the outer footpoint are of secondary 
importance to the 
process. We cannot rule out additional contributions from unavoidable 
numerical diffusivity allowing plasma to move across field lines; however, 
our analysis in Sect. \ref{sect:simple-model} did not reveal a very large 
contribution on average compared to the other terms in the induction equation.

The above interpretation shares some similarity with the "fallen flux tube" 
concept of \citet{Wentzel:1992} combined with the convective driving of 
outflows described here and previously by \citet{Scharmer:etal:2008}.

\begin{figure*}
  \centering 
  \resizebox{16.5cm}{!}{\includegraphics{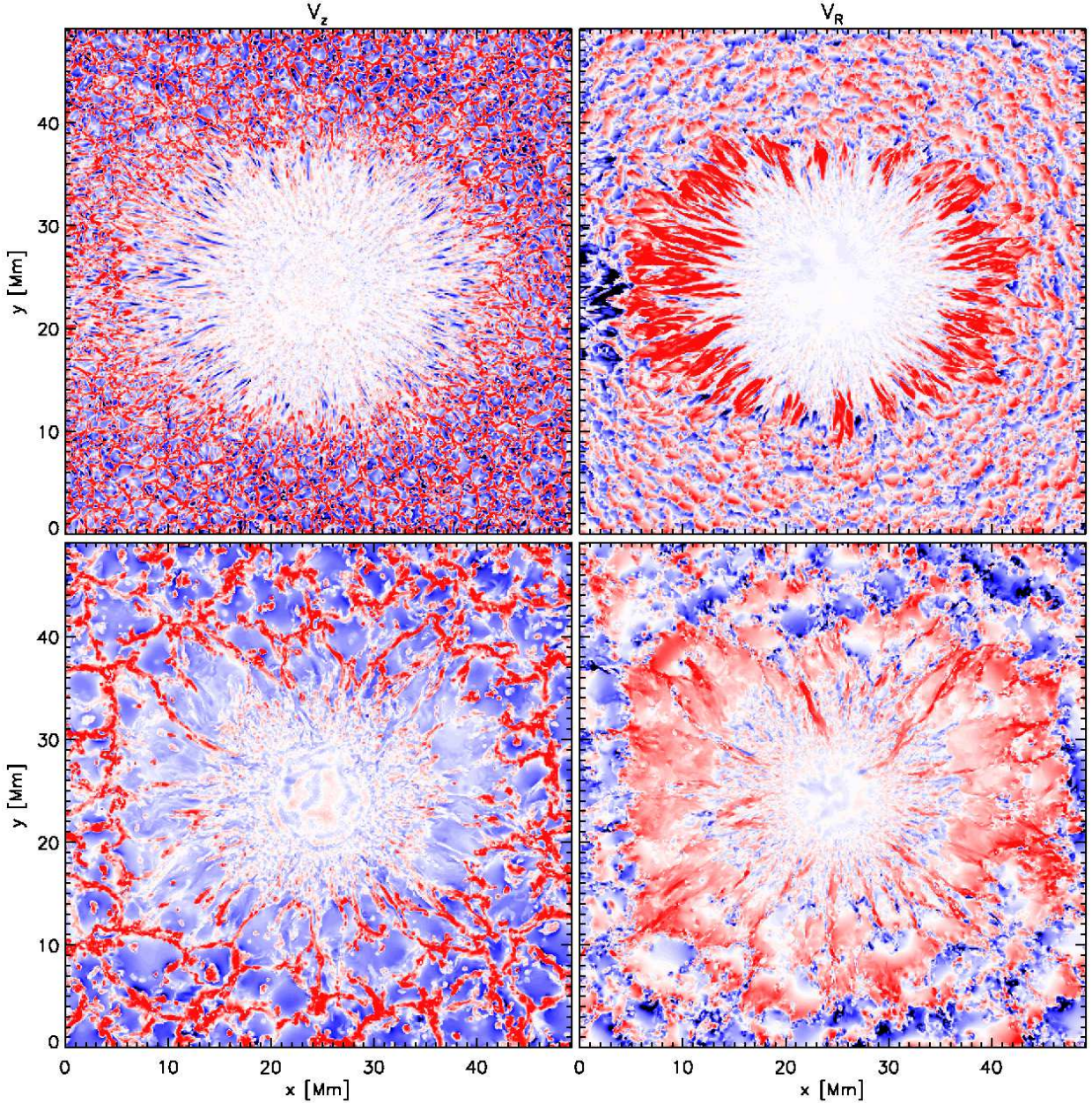}}
  \caption{Vertical (left) and radial (right) flow velocity at the 
    $\tau=1$ level (top) and $3.4$ Mm depth (bottom). At the $\tau=1$ level
    ($3.4$ Mm depth) the vertical flow velocity is saturated at 
    $3\,\mbox{km\,s}^{-1}$ ($0.8\,\mbox{km\,s}^{-1}$), the radial one at 
    $6\,\mbox{km\,s}^{-1}$ ($0.8\,\mbox{km\,s}^{-1}$). 
    While vertical motions in the penumbra are 
    subdued compared to the surrounding granulation, horizontal flows clearly
    stand out. Horizontal outflows are strongest in the x-direction, where the
    nearby opposite polarity spots impose a more horizontal field.
    In contrast to the flow 
    field in the photosphere radial motions do not stand as much over the 
    surrounding convection cells and are also much less dependent on the 
    position (x vs. y direction). The presence of the sunspot leads to 
    a ring-like arrangement of convection cells around the spot. In
    contrast to the more or less randomly arranged convection cells at larger 
    distance from the spot this preferred arrangement leads to the generation of
    mean flows with an amplitude comparable to convective flows at the given
    height.}
  \label{fig:vr_vz_tau}
\end{figure*}

\subsection{Implications for simplified models}
We address here only models that include driving processes for the
Evershed flow, a more general discussion is presented in Sect. 
\ref{sect:concl}.

The picture presented above shows substantial differences to stationary siphon
flow models that have been proposed to explain the Evershed flow. Those models
assume that processes related to turbulent pumping near the outer edge 
of the penumbra \citep{Montesinos:Thomas:1997,Brummell:etal:2008} hold field 
lines down and establish the field line connectivity that allows then for 
siphon flows due to pressure differences between the inner and outer 
footpoint. We see stronger evidence in our simulation for a flow that is
driven from within the penumbra regardless of the initial field line 
connectivity, even though siphon-like flow channels can result from this
process in the outer penumbra (see, e.g., filament 5 in Fig. 
\ref{fig:field_lines}). Since we find fast outflows along horizontal
stretches of field lines regardless of their field line connectivity we 
conjecture that siphon-like flow channels in the outer penumbra are more a 
consequence of the fast outflow than its cause. 

This does not rule out additional contributions from processes as described
by \citet{Montesinos:Thomas:1997} in the outer penumbra. The filament 6 we 
highlighted in Fig. \ref{fig:field_lines} is a potential example for a 
siphon flow related to turbulent pumping near the edge of the penumbra. The 
footpoint in the outer penumbra is caused by a strong downdraft leading to
an U-loop of field lines (see the upward returning flux in the left
corner of the indicated sub domain). The outer footpoint of filament 6 has as 
a consequence fast downflows and low pressure (on average more than
$10^5 {\rm dyne}{\rm cm}^{-2}$ lower than the regions most of the field lines
connect to further inward). We see a bundle of arch like field lines extending
a few $100$~km above $\tau=1$ and having outflow velocities mostly in the 
$2-4\,\mbox{km\,s}^{-1}$ 
range (a few faster flows are found in the proximity of the footpoint),
which is consistent with the predictions of most stationary siphon models. 
However, the flow velocities fall short of those present in filaments 2-5,
where fast outflows are confined to almost horizontal stretches 
of the field lines in the deep photosphere over lengths of several Mm. When 
there are higher reaching arches present such as in filament 3 and 4, flow 
speeds are declining toward the highpoint, contrary to predictions from 
stationary siphon flow models. 

The concentration of driving 
forces to a very narrow boundary layer beneath $\tau=1$ is not compatible 
with the acceleration of fluid along several Mm wide arches of field lines 
extending mostly above $\tau=1$. 

Fast outflows in the deep photosphere are a natural outcome of the moving 
flux tube model \citep{Schlichenmaier:etal:1998a, 
Schlichenmaier:etal:1998b} in which processes similar to convective
overshoot limit the vertical rise of plasma. The fast outflows found there
have been attributed in part to a hot upflow near the inner footpoint that
is magnetically deflected outward and in part to horizontal pressure gradients
resulting from radiative cooling. The former has some similarity to the
magnetic "deflection" we see in our magnetoconvection simulation leading to
a process limited to a very narrow height range near $\tau=1$, but we do not 
see significant contributions from horizontal pressure gradients in the
radial direction. The primary difference is that in the simplified flux tube
picture there is only one inner footpoint present for the filament, while the
driving process seen in our numerical simulation is approximately  
translation invariant along the filament, i.e. the entire filament is 
essentially a "footpoint" in which this process accelerates fluid.
The translation invariance also implies small contributions from horizontal 
pressure driving in the radial direction. 

\section{Deep flow component}
\begin{figure}
  \centering 
  \resizebox{\hsize}{!}{\includegraphics{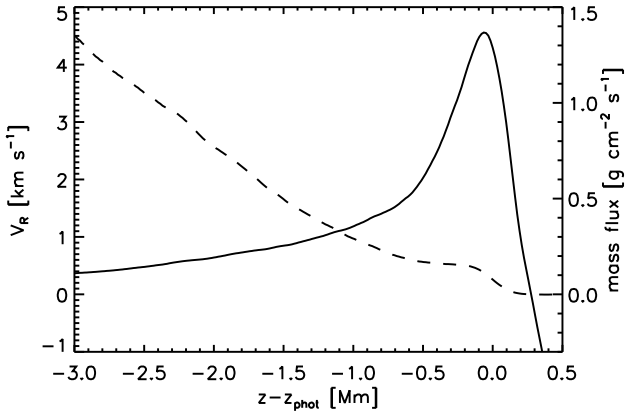}}
  \caption{Vertical velocity and mass flux profile take at $R=18$ Mm
    in the outer penumbra (cf. Fig. \ref{fig:vr_normalized}). The solid line
    shows the radial flow velocity (scale on the left) and the dashed line the
    radial mass flux $v_R \varrho$ with the scale on the right. At a depth of
    $3$ Mm the deep flow component transports about $10$ times the mass of 
    the Evershed flow due to the strong increase of mass density with depth.}
  \label{fig:mass_flux_moat}
\end{figure}  
\label{sect:deep-flow}
In the previous section we focused on the driving of flows in the uppermost 
few $100$~km of the penumbra. While these flows can be primarily explained 
through strongly enhanced Lorentz force driving, there is no strong contribution
present in more than $400$~km depth (cf. Fig. \ref{fig:energetics_center_p2}). 
The physical origin of this deeper reaching outflow becomes evident from 
Fig. \ref{fig:vr_vz_tau}. At the $\tau=1$ surface horizontal outflows 
stand out compared to typical granular flows at 
that height level, while vertical motions in the penumbra are subdued.
Furthermore, strong outflows are found preferentially along the x-direction
where the nearby opposite polarity spots impose a more horizontal magnetic
field. In about $3.4$ Mm depth vertical and
horizontal flows do not stand out in terms of amplitude, but rather in terms
of the overall flow structure. In the periphery of the sunspot convection cells
are arranged in a ring-like pattern in contrast to the random arrangement 
further away. This preferred arrangement leads to the appearance of large-scale
mean flows (outflows away from the sunspot), which have amplitudes comparable
to typical horizontal convective flows at the same depth, i.e. the flow
amplitude should be a certain fraction of the rms velocity rather independent
of depth as indicated in Fig. \ref{fig:vr_normalized}. While the appearance
of an approximately axisymmetric mean flow is a consequence of geometric 
arrangement, the preferred outflow direction requires additional ingredients.
A pure arrangement of convection cells in a ring-like fashion should lead to a
pair of convection rolls, generating an inflow close to the spot and an
outflow further out. A preference for the outflow can be a consequence of the
presence of strong magnetic field in the center, which inhibits motions 
converging toward the spot but has less influence on the diverging motions
further out. In addition large-scale pressure gradients can lead to the 
preference of outflows. We see in this simulation a combination of both.
According to Fig. \ref{fig:energetics_center_p2}a) radial flows
in the center penumbra are driven primarily by pressure forces in more than
$300$~km depth. We see a preference for pressure driving of outflows, 
only a weak asymmetry is introduced by the Lorentz force, which opposes 
inflows more strongly than outflows. Note that most of 
the pressure driving originates from the region in between $R=16$ and 
$R=18$~Mm, where the deep flow component gains speed.

While the outflow close to $\tau=1$ clearly dominates
in terms of flow velocity, the deep reaching flow component transports
significantly more mass than the shallow Evershed flow. In Fig. 
\ref{fig:mass_flux_moat}, we compare vertical profiles of $\langle v_R\rangle$ 
and $\langle v_R \varrho\rangle$ at the position $R=18$ Mm in the outer
penumbra. At a depth
of $3$ Mm the mass flux is about $10$ times larger than the the mass flux of 
the Evershed flow in the photosphere.
It is very likely that the deep flow component is related to large-scale moat
flows observed around sunspots in the photosphere as well as deeper layers
through helioseismology (see, e.g., \citet{Gizon:Birch:2005:lrsp} for a recent 
review and references therein). To clearly
establish this relationship we need however a numerical simulation covering
a substantially longer time span as well as depth range, which is beyond this 
investigation. 

\section{Conclusions}
\label{sect:concl}
We presented a detailed analysis of the recent numerical sunspot simulation by
\citet{Rempel:etal:Science}. Our investigation focused on properties of 
penumbral fine structure near the $\tau=1$ level as well as the physical 
mechanisms behind the driving of large-scale outflows in sunspots at 
photospheric levels and beneath. Our main conclusions can be summarized as 
follows:
\begin{enumerate}
  \item[$\bullet$] We find penumbral fine structure near $\tau=1$, which
    is compatible with the observationally inferred picture of an 
    interlocking comb structure with fast $>8\,\mbox{km\,s}^{-1}$ Evershed 
    flows along almost horizontal stretches of magnetic field.
  \item[$\bullet$] Correlations between radial flow velocity, intensity, 
    and field strength at the $\tau=1$ level show a good qualitative 
    agreement with recent observations and are a direct consequence of 
    convective energy transport in a sunspot penumbra.
  \item[$\bullet$] The net contribution from large-scale flows
    to the mass and convective energy flux in the penumbra is found to
    be about $12-13\%$. Local contributions in the inner penumbra
    can reach $50\%$. Maintaining the penumbral brightness of $0.7 L_{\odot}$
    requires about $1\,\mbox{km\,s}^{-1}$ vertical rms velocity at the
    $\tau=1$ level.
  \item[$\bullet$] We find in the sunspot penumbra two flow components, which we
    separate according to their scaling with respect to the convective 
    rms velocity outside the penumbra ($v_{\rm rms}^0$). While the deep 
    component (more than $500$ km beneath the photosphere) has an almost
    constant ratio to $v_{\rm rms}^0$ of about $50\%$ rather independent of 
    depth, the shallow component (uppermost $500$ km)  shows an increase
    toward the surface steeper than $v_{\rm rms}^0$. While the latter is 
    essentially the Evershed flow, the former is likely related to a deep 
    reaching moat flow component.
  \item[$\bullet$] The near surface flow component is almost exclusively 
    driven through the horizontal component of the Lorentz force along 
    filaments. 
    The energy for maintaining this flow is provided by vertical pressure
    forces in upflow regions, which are effectively deflected horizontally 
    by an inclined magnetic field.
  \item[$\bullet$] The Evershed flow is driven in a thin boundary layer 
    beneath $\tau=1$. Essential ingredient is a strong vertical increase of 
    $B_R$ beneath $\tau=1$ combined with a moderate vertical average field 
    strength
    of a few $100$ G. The Evershed flow is strongly magnetized and 
      reaches a peak velocity of about $\vert B_R\vert/\sqrt{4\pi\varrho}$ 
      at $\tau=1$.
  \item[$\bullet$] Upward advection of momentum and magnetic field by 
    overturning convective motions in the penumbra is crucial for 
    maintaining the conditions under which a quasi-stationary Evershed flow
    can be driven.
  \item[$\bullet$] The deep reaching flow component results from a preferred
    geometric alignment of convection cells in the periphery of the sunspot.
    Asymmetries in pressure and Lorentz forces lead to a dominance of the
    outflow component. The flow amplitude is about $50\%$  of the convective
    rms velocity, rather independent of depth. 
  \item[$\bullet$] Flow channels cannot be easily identified with 
      magnetic flux tubes. The field line connectivity is changing between 
      inner and outer penumbra and we see no compelling evidence that the 
      field line connectivity plays a major role in determining the Evershed 
      flow speed; on the contrary, field line connectivity is established
      primarily as a consequence of the outflow.
\end{enumerate}  

A variety of simplified as well as magnetoconvective models for the penumbra
have been discussed to explain the Evershed effect. The majority of the 
simplified models that have been used to describe the acceleration of 
horizontal flows in the penumbra is based on stationary or dynamic flux 
tube models. 

We see strong limitations for the applicability of the thin flux tube 
approximation in the context of penumbral flow channels. The continuous
mass exchange along flow channels due to overturning convective motions 
is not part of the flux tube picture (it is essentially orthogonal to the
assumptions), but found to be substantial in the presented numerical
simulation. In addition the changing field line connectivity along the flow 
channels from inner to outer penumbra does not allow for an easy 
identification with a flux tube, at least 
not for the whole length of it. It is in general difficult to find meaningful
compact footpoints that are representative of the field in inner and outer
penumbra at the same time. 

We see limitations for the applicability of 
stationary flux tube models such as \citet{Meyer:Schmidt:1968, 
Thomas:1988,Degenhardt:1989,Degenhardt:1991,Thomas:Montesinos:1993,
Montesinos:Thomas:1997} 
as explanation for the
flows in our simulation. Both, field line connectivity and the causality
between outflows and field line connectivity point toward processes in
which the conditions in the outer footpoint (if it exists at all) are of
minor influence on the flow pattern and outflows are mostly driven
by convective motions within the penumbra. In addition fast outflows are 
found preferentially in the deep photosphere along almost horizontal 
stretches of field lines regardless of their connectivity, which is different 
from the situation described in most stationary siphon models to date.
This does not rule out additional contributions in the outer penumbra from
processes similar to those described in \citet{Montesinos:Thomas:1997}, where 
turbulent pumping by convective motions at the periphery of the penumbra 
plays a crucial role in establishing the field geometry. Filament 6 in
Fig. \ref{fig:field_lines} is one possible example for such a configuration. 
Note that our simulation might not be fully representative of a "typical" 
outer penumbra due to the setup with a nearby opposite polarity spot, 
even though, the combination of strong horizontal field in between both 
sunspots with a strong downflow lane due to the converging Evershed flows 
from both sides has a tendency to enhance submergence of field by convective 
motions. 
    
The moving flux tube model of \citet{Schlichenmaier:etal:1998a, 
Schlichenmaier:etal:1998b} naturally produces outflows located in the 
deep photosphere and similarly to the situation in our simulation most driving
is focused on the inner footpoint. The acceleration of outflows is attributed
to a combination of "deflection" of hot upflows and horizontal pressure 
gradients resulting from radiative cooling. In contrast to moving flux
tube models we see in our simulation strong limits on the overall role 
horizontal pressure driving plays for the acceleration of plasma in the 
radial direction, while the magnetic "deflection" of hot upflows is not
restricted to the inner footpoint but found everywhere throughout the
penumbra.

Recently \citet{Thomas:2010} and also Priest (2010, private
communication) have suggested that flows in the penumbra
could be described in terms of “dynamical” siphon models,
in which pressure gradients are produced by MHD processes such as
magnetoconvection and these drive time-dependent flows along the
magnetic field, which then reacts by Lorentz forces to the presence of
the flow: in this view, field line connectivity  plays only a
secondary role and is more a consequence than a cause.
While describing part of the picture, however, one also needs to
understand the fluid motions that are responsible for the
filamentation and most of the energy transport.
Furthermore, we showed that the magnetic driving of outflows originates from
narrow boundary layers that form beneath $\tau=1$ in regions where
convective upflows are present. These regions arise as a consequence of
magnetoconvection and are not captured by flux tube models that do not
include a filament sub-structure and overturning convection.

\citet{Galloway:1975} presented a phenomenological model based on the 
"roll-convection" picture introduced by \citet{Danielson:1961}, which
explains the Evershed flow as a consequence of unbalanced horizontal Lorentz
force components: while the Lorentz force balances the gas
pressure deficit of the sunspot on average, the filamentation of the penumbra 
implies a strong azimuthal variation and therefore a violation of this balance 
locally that is responsible for driving the Evershed flow. It is argued 
further that the Evershed flow is located in dark filaments, where downward 
directed motions concentrate magnetic field and lead in return to an above 
average Lorentz force. While our magnetoconvection simulation certainly 
produces a Lorentz force that is strongly varying in the 
azimuthal direction, our results disagree with \citet{Galloway:1975} in 
detail as we find that most of the Evershed flow is driven in upflow regions. 

Several other penumbra models focus on the fine structure of the penumbra,
without necessarily addressing the driving of large-scale outflows. One
of these models which has gotten a lot of attention recently is the "gappy"
penumbra model by \citet{Spruit:Scharmer:2006,Scharmer:Spruit:2006}. In its
original form this model proposed essentially field free upflow plumes
embedded in an inclined background field. While this structure certainly
captures the essence of the convective picture seen in the numerical 
simulations presented here, there has been a lot of discussion of how
field free these "gaps" really are and to which extent the field strength
seen in present simulations is affected by numerical diffusion 
\citep[see, e.g., discussion in][]{Nordlund:Scharmer:2010}. The physical
explanation of the Evershed flow presented here implies the presence
of strong Lorentz forces in the uppermost few $100$ km beneath the $\tau=1$
level in the penumbra, which requires the presence of strong
$1-2$ kG magnetic field. In that sense our results
are incompatible with models that predict field free gaps at photospheric 
levels! Furthermore we do not see evidence that the field structure is heavily
influenced by numerical dissipation as the analysis presented in the bottom
panels of Fig. \ref{fig:filament_avr} reveals. Also it remains very
controversial whether the wealth of spectropolarimetric observations could be
explained by an essentially unmagnetized Evershed flow 
\citep[see, e.g.,][]{Thomas:2010}. Nevertheless, the results presented here
indicate an almost Salomonian solution to this discussion: Strong field
is confined to a narrow boundary layer just beneath $\tau=1$, while
further down the field strength is substantially reduced compared to the 
ambient plasma (but still of the order of $1$~kG). This scenario was also
brought forward as a possible solution for this problem in a recent review 
by \citet{Scharmer:2009}.

Recently a variety of different magnetoconvection simulations with
radiative transfer such as \citet{Heinemann:etal:2007,Rempel:etal:2009,
Kitiashvili:etal:2009} have been used to model the penumbra. The
models by \citet{Heinemann:etal:2007} and \citet{Rempel:etal:2009} focused
primarily on the transition from umbra toward inner penumbra, which 
corresponds roughly to the innermost edge of the region we analyzed in this
investigation. The energy conversion terms for that region are
displayed in Fig. \ref{fig:energetics_inner_p} and do not show (except for
the overall amplitude) a fundamental difference to 
Fig. \ref{fig:energetics_center_p}. From this we conclude that the driving
mechanisms for horizontal outflows in \citet{Heinemann:etal:2007} and 
\citet{Rempel:etal:2009} is essentially the same as discussed here. The
overall picture we described in Sect. \ref{sect:filaments} is similar to
\citet{Scharmer:etal:2008}.
\citet{Kitiashvili:etal:2009} studied in an idealized setup the 
influence of field strength and inclination on large-scale flows and found 
a strong dependence of outflow speeds on the average inclination and field
strength, which is consistent with the mechanism explained here. They also
reported on temporal variations of the Evershed flow speed on timescales
of $15$ to $40$ minutes.

Our analysis essentially reinforces conclusions of 
\citet{Scharmer:etal:2008,Rempel:etal:Science} that the penumbra is anisotropic
magnetoconvection and that the Evershed flow can be understood as convective 
flow component in the direction of the magnetic field. The similarity between 
plage region and penumbra with respect to the different terms in the
kinetic energy equation (Fig. \ref{fig:energetics_cmp}) is quite astonishing, 
a comparison between the depth profiles of rms velocities points toward
anisotropy as the main difference (see \citet{Rempel:etal:Science},
supporting online material). Nevertheless, there are also notable differences
which clearly differentiate the Evershed flow from horizontal flows 
in typical convection. While the latter is entirely pressure driven, the 
Evershed flow is almost completely Lorentz force driven. Only flows that 
turn over laterally in penumbral filaments remain pressure driven. 
In addition pressure/buoyancy driving takes place primarily in upflow regions, 
in contrast to field free convection that is driven by top heavy downflow
regions. An interesting new aspect pointed out in this paper is the confinement
of the underlying driving mechanism to very narrow boundary layers that 
exist just beneath $\tau=1$.
 
For clarification we want point out that the driving of the Evershed flow
is achieved through the radial component of the Lorentz force while the total 
work done by Lorentz forces remains negative, i.e. the net effect is a sink 
for kinetic energy. The overall underlying energy source is convective 
instability, which enters the kinetic energy balance through pressure/buoyancy
driving. The Lorentz force facilitates the energy exchange between the
pressure driving in the vertical direction and the horizontal Evershed flow 
acceleration. A necessary condition for the latter is the shift of 
pressure/buoyancy driving from downflow to upflow regions in the penumbra 
we described above.

In addition to the mechanism leading to the fast Evershed flow in the upper 
most few $100$ km of a sunspot penumbra we have also identified a mechanism 
leading to the formation of a larger scale outflow in deeper layers. In 
contrast to the Evershed flow the deeper flow scales proportional to the
convective rms velocity (outside the sunspot), the dominant radial outflow 
reaches typically 
amplitudes $\sim 0.5\,v_{\rm rms}$. The main reason for this flow cell is a 
preferred circular alignment of convection cells surrounding the sunspot.
As a consequence the azimuthal average over this ring-like pattern 
of convection cells does not vanish and leads to mean flow speeds scaling 
proportional to $v_{\rm rms}$. A preference for the outflow results from a 
combination of pressure and Lorentz forces. While this flow does not stand 
out in terms of flow velocity as the Evershed flow, the radial mass flux is
substantially larger, which make a connection with the large-scale moat flows
observed around sunspots likely. In this simulation we do not see evidence for
a converging collar flow that was found previously in 2D axisymmetric 
simulations \citep{Hurlburt:Rucklidge:2000,Botha:etal:2006,Botha:etal:2008}.
In a future publication we will investigate the subsurface structure 
of this flow component in deeper domains and evolution over timescales 
longer than those covered by the numerical simulation presented here.

The deep flow component described in Sect. \ref{sect:deep-flow} should be
in principle observable through local helioseismology. The clear prediction
is here an outflow of plasma with an amplitude of about $50\%$ of the
convective rms velocity reaching downward several Mm beneath the penumbra.
This result is in contradiction with some recent helioseismic inversions such 
as \citet{Zhao:etal:2001,Zhao:etal:2010}, which point toward an inflow in a 
depth range from $1.5$ to $5$ Mm. One the other hand \citet{Gizon:etal:2009}
reported on an outflow over the uppermost $5$~Mm.

While most of the processes responsible for driving the Evershed flow
are located beneath the $\tau=1$ level,
there are nevertheless several aspects of the magnetoconvective penumbra
model presented here that can be constrained through observations. As presented 
in Sect. \ref{sect:phot} most of the energy is transported in the
penumbra by laterally overturning convective motions. We find a very
tight relationship between intensity and vertical rms velocity of the
form $I \propto \sqrt{v_{z\,rms}(\tau=1)}$. From this follows that the 
vertical rms velocity at $\tau=1$ in the penumbra with $I\approx 0.7 I_{\odot}$ 
should be about half of the value found in the quiet sun, i.e. about 1 
$\mbox{km\,s}^{-1}$. This value is consistent with the recent findings of 
\citet{Franz:Schlichenmaier:2009}, who computed from {\em Hinode} observations
velocity distributions functions for both quiet Sun and penumbra (see
Fig.3 in their paper). The half width at half maximum of the vertical
velocity distribution function for the penumbra is about 
$500\,\mbox{m\,s}^{-1}$, while the same analysis results in 
$1\,\mbox{km\,s}^{-1}$ for the quiet Sun, i.e. the latter
falls short by about a factor of $2$ compared to the value we find for the 
quiet Sun at $\tau=1$ (due to a combination of limited observational 
resolution as well as the sharp decline of $v_z$ above $\tau=1$). 
If we assume that the same shortfall applies also to the penumbra, 
the vertical velocity structure reported in \citet{Franz:Schlichenmaier:2009} 
is at least in a statistical sense fully consistent with the amount of
overturning convection we see in the numerical simulation presented here. 
Other consequences of the magnetoconvective model are the sign
changes in the $I - v_R$ and $B - v_R$ correlations presented in 
Fig. \ref{fig:corr}. They are consistent with the analysis of {\em Hinode} data 
presented by \citet{Ichimoto:etal:2007} (see Fig. 3 therein). A 
positive $B - v_R$ correlation in the outer penumbra was suggested by 
\citet{Tritschler:etal:2007} and \citet{Ichimoto:etal:2008} based on 
observations of the net circular polarization (NCP) at different viewing 
angles. As explained in Sect. \ref{sect:simple-model}
the observable consequences of the Evershed flow driving mechanism are a
moderate increase of $B_R$ and a steep increase of $v_R$ 
toward $\tau=1$. $v_z$ shows a very steep gradient in the deep photosphere. 
The peak velocity of the Evershed flow in the deep photosphere should be
around $\vert B_R\vert/\sqrt{4\pi\varrho}$.

The fact that the simulated Evershed flow is a deep photospheric flow is
a direct consequence of its convective origin. On the observational
side the depth dependence of the Evershed flow is debated. While the 
investigations by \citet{Rimmele:1995} and \citet{Stanchfield:etal:1997} point 
toward flows in elevated flow channels, recent work by  
\citet{Schlichenmaier:etal:2004}, \citet{BellotRubio:etal:2006}, and
\citet{Borrero:etal:2008} is in support of a flow in the deep photosphere 
declining with height. Another point heavily
debated is the presence or absence of overturning convection in the 
penumbra. Support for overturning convection is found by 
\citet{Ichimoto:etal:2007:sc,Zakharov:etal:2008,Rimmele:2008,
Bharti:etal:2010}, while
\citet{BellotRubio:etal:2005,Ichimoto:etal:2007,Franz:Schlichenmaier:2009,
BellotRubio:etal:2010} see primarily support for Evershed flow related 
upflows in the inner and downflows in the outer penumbra -- a flow pattern
that accounts in our model only for a small fraction of the unsigned
mass and energy flux integrated over the penumbra. It appears that overcoming
the discrepancy between the presence of overturning convection in MHD 
simulations and the lack of evidence in many high resolution observations is 
one of the biggest challenges both numerical models and observations will
face in the future. It is unlikely that the absence of overturning convection
is the solution to this discrepancy; a brightness of $0.7 I_{\odot}$ or more
requires overturning mass flux at a level not much less than granulation.
An other possible solution could be related to thin boundary layers, which are
indicated but not well resolved in the simulation presented here. If quantities
such as flow velocities and magnetic field change dramatically over short
distances, moving a $\tau$-surface by a distance comparable to our grid 
spacing can make a dramatic differences for the visibility of such feature.
This clearly indicates that the simulation presented here can only be 
considered as a first step in that direction.

A convergence study of the properties highlighted in this investigation 
covering the resolution range from $96\times 32$ to $16\times 12$~km resolution
(horizontal $\times$ vertical) is in progress. A preliminary analysis shows
that most magnetoconvective properties of the penumbra are robust 
(qualitative agreement over the whole range investigated, quantitative 
agreement from $48\times 24$~km resolution upward), while the photospheric 
appearance of sunspot fine structure improves substantially with resolution.
The currently highest resolution case is presented in \citet{Rempel:2010:IAU}.

\acknowledgements
The National Center for Atmospheric Research is sponsored by the National 
Science Foundation. Computing resources were provided by NCAR's 
Computational and Information Systems Laboratory (CISL). M. Rempel is grateful
to Manfred Sch{\"u}ssler, Alfred De Wijn, Michael Kn{\"o}lker, Keith MacGregor,
and Eric Priest for helpful comments on the manuscript. M. Rempel also thanks
the referee of this paper, J.H. Thomas, for providing feedback that 
substantially improved the presentation.

\bibliographystyle{natbib/apj}
\bibliography{natbib/apj-jour,natbib/papref_m}

\begin{thebibliography}{74}
\expandafter\ifx\csname natexlab\endcsname\relax\def\natexlab#1{#1}\fi

\bibitem[{{Bellot Rubio} {et~al.}(2005){Bellot Rubio}, {Langhans}, \&
  {Schlichenmaier}}]{BellotRubio:etal:2005}
{Bellot Rubio}, L.~R., {Langhans}, K., \& {Schlichenmaier}, R. 2005, \aap, 443,
  L7

\bibitem[{{Bellot Rubio} {et~al.}(2010){Bellot Rubio}, {Schlichenmaier}, \&
  {Langhans}}]{BellotRubio:etal:2010}
{Bellot Rubio}, L.~R., {Schlichenmaier}, R., \& {Langhans}, K. 2010, \apj, 725,
  11

\bibitem[{{Bellot Rubio} {et~al.}(2006){Bellot Rubio}, {Schlichenmaier}, \&
  {Tritschler}}]{BellotRubio:etal:2006}
{Bellot Rubio}, L.~R., {Schlichenmaier}, R., \& {Tritschler}, A. 2006, \aap,
  453, 1117

\bibitem[{{Bharti} {et~al.}(2010){Bharti}, {Solanki}, \&
  {Hirzberger}}]{Bharti:etal:2010}
{Bharti}, L., {Solanki}, S.~K., \& {Hirzberger}, J. 2010, \apjl, 722, L194

\bibitem[{{Borrero} {et~al.}(2008){Borrero}, {Lites}, \&
  {Solanki}}]{Borrero:etal:2008}
{Borrero}, J.~M., {Lites}, B.~W., \& {Solanki}, S.~K. 2008, \aap, 481, L13

\bibitem[{{Botha} {et~al.}(2008){Botha}, {Busse}, {Hurlburt}, \&
  {Rucklidge}}]{Botha:etal:2008}
{Botha}, G.~J.~J., {Busse}, F.~H., {Hurlburt}, N.~E., \& {Rucklidge}, A.~M.
  2008, \mnras, 387, 1445

\bibitem[{{Botha} {et~al.}(2006){Botha}, {Rucklidge}, \&
  {Hurlburt}}]{Botha:etal:2006}
{Botha}, G.~J.~J., {Rucklidge}, A.~M., \& {Hurlburt}, N.~E. 2006, \mnras, 369,
  1611

\bibitem[{{Brummell} {et~al.}(2008){Brummell}, {Tobias}, {Thomas}, \&
  {Weiss}}]{Brummell:etal:2008}
{Brummell}, N.~H., {Tobias}, S.~M., {Thomas}, J.~H., \& {Weiss}, N.~O. 2008,
  \apj, 686, 1454

\bibitem[{{Cabrera Solana} {et~al.}(2007){Cabrera Solana}, {Bellot Rubio},
  {Beck}, \& {Del Toro Iniesta}}]{Cabrera:etal:2007}
{Cabrera Solana}, D., {Bellot Rubio}, L.~R., {Beck}, C., \& {Del Toro Iniesta},
  J.~C. 2007, \aap, 475, 1067

\bibitem[{{Clyne} {et~al.}(2007){Clyne}, {Mininni}, {Norton}, \&
  {Rast}}]{Clyne:etal:2007}
{Clyne}, J., {Mininni}, P., {Norton}, A., \& {Rast}, M. 2007, New Journal of
  Physics, 9, 301

\bibitem[{{Clyne} \& {Rast}(2005)}]{Clyne:Rast:2005}
{Clyne}, J. \& {Rast}, M. 2005, in Society of Photo-Optical Instrumentation
  Engineers (SPIE) Conference Series, Vol. 5669, Society of Photo-Optical
  Instrumentation Engineers (SPIE) Conference Series, ed. {R.~F.~Erbacher,
  J.~C.~Roberts, M.~T.~Gr{\"o}hn, \& K.~B{\"o}rner }, 284--294

\bibitem[{{Danielson}(1961)}]{Danielson:1961}
{Danielson}, R.~E. 1961, \apj, 134, 289

\bibitem[{{Degenhardt}(1989)}]{Degenhardt:1989}
{Degenhardt}, D. 1989, \aap, 222, 297

\bibitem[{{Degenhardt}(1991)}]{Degenhardt:1991}
---. 1991, \aap, 248, 637

\bibitem[{{del Toro Iniesta} {et~al.}(2001){del Toro Iniesta}, {Bellot Rubio},
  \& {Collados}}]{DelToro:etal:2001}
{del Toro Iniesta}, J.~C., {Bellot Rubio}, L.~R., \& {Collados}, M. 2001,
  \apjl, 549, L139

\bibitem[{{Dialetis} {et~al.}(1985){Dialetis}, {Mein}, \&
  {Alissandrakis}}]{Dialetis:etal:1985}
{Dialetis}, D., {Mein}, P., \& {Alissandrakis}, C.~E. 1985, \aap, 147, 93

\bibitem[{{Evershed}(1909)}]{Evershed:1909}
{Evershed}, J. 1909, \mnras, 69, 454

\bibitem[{{Franz} \& {Schlichenmaier}(2009)}]{Franz:Schlichenmaier:2009}
{Franz}, M. \& {Schlichenmaier}, R. 2009, \aap, 508, 1453

\bibitem[{{Galloway}(1975)}]{Galloway:1975}
{Galloway}, D.~J. 1975, \solphys, 44, 409

\bibitem[{{Gizon} \& {Birch}(2005)}]{Gizon:Birch:2005:lrsp}
{Gizon}, L. \& {Birch}, A.~C. 2005, Living Reviews in Solar Physics, 2, 6

\bibitem[{{Gizon} {et~al.}(2009){Gizon}, {Schunker}, {Baldner}, {Basu},
  {Birch}, {Bogart}, {Braun}, {Cameron}, {Duvall}, {Hanasoge}, {Jackiewicz},
  {Roth}, {Stahn}, {Thompson}, \& {Zharkov}}]{Gizon:etal:2009}
{Gizon}, L., {Schunker}, H., {Baldner}, C.~S., {Basu}, S., {Birch}, A.~C.,
  {Bogart}, R.~S., {Braun}, D.~C., {Cameron}, R., {Duvall}, T.~L., {Hanasoge},
  S.~M., {Jackiewicz}, J., {Roth}, M., {Stahn}, T., {Thompson}, M.~J., \&
  {Zharkov}, S. 2009, \ssr, 144, 249

\bibitem[{{Grosser}(1991)}]{Grosser:1991}
{Grosser}, H. 1991, Zur Entstehung der Penumbra-Filamentierung von
  Sonnenflecken durch die Wirkung von Konvektionsrollen, PhD thesis,
  Universit\"at G\"ottingen

\bibitem[{{Heinemann} {et~al.}(2007){Heinemann}, {Nordlund}, {Scharmer}, \&
  {Spruit}}]{Heinemann:etal:2007}
{Heinemann}, T., {Nordlund}, {\AA}., {Scharmer}, G.~B., \& {Spruit}, H.~C.
  2007, \apj, 669, 1390

\bibitem[{{Hurlburt} {et~al.}(1996){Hurlburt}, {Matthews}, \&
  {Proctor}}]{Hurlburt:etal:1996}
{Hurlburt}, N.~E., {Matthews}, P.~C., \& {Proctor}, M.~R.~E. 1996, \apj, 457,
  933

\bibitem[{{Hurlburt} {et~al.}(2000){Hurlburt}, {Matthews}, \&
  {Rucklidge}}]{Hurlburt:etal:2000}
{Hurlburt}, N.~E., {Matthews}, P.~C., \& {Rucklidge}, A.~M. 2000, \solphys,
  192, 109

\bibitem[{{Hurlburt} \& {Rucklidge}(2000)}]{Hurlburt:Rucklidge:2000}
{Hurlburt}, N.~E. \& {Rucklidge}, A.~M. 2000, \mnras, 314, 793

\bibitem[{{Ichimoto} {et~al.}(2007{\natexlab{a}}){Ichimoto}, {Shine}, {Lites},
  {Kubo}, {Shimizu}, {Suematsu}, {Tsuneta}, {Katsukawa}, {Tarbell}, {Title},
  {Nagata}, {Yokoyama}, \& {Shimojo}}]{Ichimoto:etal:2007}
{Ichimoto}, K., {Shine}, R.~A., {Lites}, B., {Kubo}, M., {Shimizu}, T.,
  {Suematsu}, Y., {Tsuneta}, S., {Katsukawa}, Y., {Tarbell}, T.~D., {Title},
  A.~M., {Nagata}, S., {Yokoyama}, T., \& {Shimojo}, M. 2007{\natexlab{a}},
  \pasj, 59, 593

\bibitem[{{Ichimoto} {et~al.}(2007{\natexlab{b}}){Ichimoto}, {Suematsu},
  {Tsuneta}, {Katsukawa}, {Shimizu}, {Shine}, {Tarbell}, {Title}, {Lites},
  {Kubo}, \& {Nagata}}]{Ichimoto:etal:2007:sc}
{Ichimoto}, K., {Suematsu}, Y., {Tsuneta}, S., {Katsukawa}, Y., {Shimizu}, T.,
  {Shine}, R.~A., {Tarbell}, T.~D., {Title}, A.~M., {Lites}, B.~W., {Kubo}, M.,
  \& {Nagata}, S. 2007{\natexlab{b}}, Science, 318, 1597

\bibitem[{{Ichimoto} {et~al.}(2008){Ichimoto}, {Tsuneta}, {Suematsu},
  {Katsukawa}, {Shimizu}, {Lites}, {Kubo}, {Tarbell}, {Shine}, {Title}, \&
  {Nagata}}]{Ichimoto:etal:2008}
{Ichimoto}, K., {Tsuneta}, S., {Suematsu}, Y., {Katsukawa}, Y., {Shimizu}, T.,
  {Lites}, B.~W., {Kubo}, M., {Tarbell}, T.~D., {Shine}, R.~A., {Title}, A.~M.,
  \& {Nagata}, S. 2008, \aap, 481, L9

\bibitem[{{Kitiashvili} {et~al.}(2009){Kitiashvili}, {Kosovichev}, {Wray}, \&
  {Mansour}}]{Kitiashvili:etal:2009}
{Kitiashvili}, I.~N., {Kosovichev}, A.~G., {Wray}, A.~A., \& {Mansour}, N.~N.
  2009, \apjl, 700, L178

\bibitem[{{Langhans} {et~al.}(2007){Langhans}, {Scharmer}, {Kiselman}, \&
  {L{\"o}fdahl}}]{Langhans:etal:2007}
{Langhans}, K., {Scharmer}, G.~B., {Kiselman}, D., \& {L{\"o}fdahl}, M.~G.
  2007, \aap, 464, 763

\bibitem[{{Langhans} {et~al.}(2005){Langhans}, {Scharmer}, {Kiselman},
  {L{\"o}fdahl}, \& {Berger}}]{Langhans:etal:2005}
{Langhans}, K., {Scharmer}, G.~B., {Kiselman}, D., {L{\"o}fdahl}, M.~G., \&
  {Berger}, T.~E. 2005, \aap, 436, 1087

\bibitem[{{Meyer} \& {Schmidt}(1968)}]{Meyer:Schmidt:1968}
{Meyer}, F. \& {Schmidt}, H.~U. 1968, Zeitschrift Angewandte Mathematik und
  Mechanik, 48, 218

\bibitem[{{Montesinos} \& {Thomas}(1997)}]{Montesinos:Thomas:1997}
{Montesinos}, B. \& {Thomas}, J.~H. 1997, \nat, 390, 485

\bibitem[{{Nordlund} \& {Scharmer}(2010)}]{Nordlund:Scharmer:2010}
{Nordlund}, {\AA}. \& {Scharmer}, G.~B. 2010, in Magnetic Coupling between the
  Interior and Atmosphere of the Sun, ed. {S.~S.~Hasan \& R.~J.~Rutten},
  243--254

\bibitem[{{Ortiz} {et~al.}(2010){Ortiz}, {Bellot Rubio}, \& {Rouppe van der
  Voort}}]{Ortiz:etal:2010}
{Ortiz}, A., {Bellot Rubio}, L.~R., \& {Rouppe van der Voort}, L. 2010, \apj,
  713, 1282

\bibitem[{{Rempel}(2010)}]{Rempel:2010:IAU}
{Rempel}, M. 2010, IAU Symp. 273, arXiv:1011.0981

\bibitem[{{Rempel} {et~al.}(2009{\natexlab{a}}){Rempel}, {Sch{\"u}ssler},
  {Cameron}, \& {Kn{\"o}lker}}]{Rempel:etal:Science}
{Rempel}, M., {Sch{\"u}ssler}, M., {Cameron}, R.~H., \& {Kn{\"o}lker}, M.
  2009{\natexlab{a}}, Science, 325, 171

\bibitem[{{Rempel} {et~al.}(2009{\natexlab{b}}){Rempel}, {Sch{\"u}ssler}, \&
  {Kn{\"o}lker}}]{Rempel:etal:2009}
{Rempel}, M., {Sch{\"u}ssler}, M., \& {Kn{\"o}lker}, M. 2009{\natexlab{b}},
  \apj, 691, 640

\bibitem[{{Riethm{\"u}ller} {et~al.}(2008){Riethm{\"u}ller}, {Solanki}, \&
  {Lagg}}]{Riethmueller:etal:2008}
{Riethm{\"u}ller}, T.~L., {Solanki}, S.~K., \& {Lagg}, A. 2008, \apjl, 678,
  L157

\bibitem[{{Rimmele}(2008)}]{Rimmele:2008}
{Rimmele}, T. 2008, \apj, 672, 684

\bibitem[{{Rimmele} \& {Marino}(2006)}]{Rimmele:Marino:2006}
{Rimmele}, T. \& {Marino}, J. 2006, \apj, 646, 593

\bibitem[{{Rimmele}(1994)}]{Rimmele:1994}
{Rimmele}, T.~R. 1994, \aap, 290, 972

\bibitem[{{Rimmele}(1995)}]{Rimmele:1995}
---. 1995, \aap, 298, 260

\bibitem[{{Scharmer}(2009)}]{Scharmer:2009}
{Scharmer}, G.~B. 2009, \ssr, 144, 229

\bibitem[{{Scharmer} {et~al.}(2002){Scharmer}, {Gudiksen}, {Kiselman}, {L{\"
  o}fdahl}, \& {Rouppe van der Voort}}]{Scharmer:etal:2002}
{Scharmer}, G.~B., {Gudiksen}, B.~V., {Kiselman}, D., {L{\" o}fdahl}, M.~G., \&
  {Rouppe van der Voort}, L.~H.~M. 2002, \nat, 420, 151

\bibitem[{{Scharmer} {et~al.}(2007){Scharmer}, {Langhans}, {Kiselman}, \&
  {L{\"o}fdahl}}]{Scharmer:etal:2007}
{Scharmer}, G.~B., {Langhans}, K., {Kiselman}, D., \& {L{\"o}fdahl}, M.~G.
  2007, in Astronomical Society of the Pacific Conference Series, Vol. 369, New
  Solar Physics with Solar-B Mission, ed. {K.~Shibata, S.~Nagata, \&
  T.~Sakurai}, 71--+

\bibitem[{{Scharmer} {et~al.}(2008){Scharmer}, {Nordlund}, \&
  {Heinemann}}]{Scharmer:etal:2008}
{Scharmer}, G.~B., {Nordlund}, {\AA}., \& {Heinemann}, T. 2008, \apjl, 677,
  L149

\bibitem[{{Scharmer} \& {Spruit}(2006)}]{Scharmer:Spruit:2006}
{Scharmer}, G.~B. \& {Spruit}, H.~C. 2006, \aap, 460, 605

\bibitem[{{Schlichenmaier} {et~al.}(2005){Schlichenmaier}, {Bellot Rubio}, \&
  {Tritscher}}]{Schlichenmaier:etal:2005}
{Schlichenmaier}, R., {Bellot Rubio}, L.~R., \& {Tritscher}, A. 2005,
  Astronomische Nachrichten, 326, 301

\bibitem[{{Schlichenmaier} {et~al.}(2004){Schlichenmaier}, {Bellot Rubio}, \&
  {Tritschler}}]{Schlichenmaier:etal:2004}
{Schlichenmaier}, R., {Bellot Rubio}, L.~R., \& {Tritschler}, A. 2004, \aap,
  415, 731

\bibitem[{{Schlichenmaier} {et~al.}(1998{\natexlab{a}}){Schlichenmaier},
  {Jahn}, \& {Schmidt}}]{Schlichenmaier:etal:1998a}
{Schlichenmaier}, R., {Jahn}, K., \& {Schmidt}, H.~U. 1998{\natexlab{a}}, \apj,
  493, L121

\bibitem[{{Schlichenmaier} {et~al.}(1998{\natexlab{b}}){Schlichenmaier},
  {Jahn}, \& {Schmidt}}]{Schlichenmaier:etal:1998b}
---. 1998{\natexlab{b}}, \aap, 337, 897

\bibitem[{{Sch{\"u}ssler} \& {V{\"o}gler}(2006)}]{Schuessler:Voegler:2006}
{Sch{\"u}ssler}, M. \& {V{\"o}gler}, A. 2006, \apjl, 641, L73

\bibitem[{{Shine} {et~al.}(1994){Shine}, {Title}, {Tarbell}, {Smith}, {Frank},
  \& {Scharmer}}]{Shine:etal:1994}
{Shine}, R.~A., {Title}, A.~M., {Tarbell}, T.~D., {Smith}, K., {Frank}, Z.~A.,
  \& {Scharmer}, G. 1994, \apj, 430, 413

\bibitem[{{Sobotka} \& {Jur{\v c}{\'a}k}(2009)}]{Sobotka:Jurcak:2009}
{Sobotka}, M. \& {Jur{\v c}{\'a}k}, J. 2009, \apj, 694, 1080

\bibitem[{{Socas-Navarro} {et~al.}(2004){Socas-Navarro}, {Pillet}, {Sobotka},
  \& {V{\' a}zquez}}]{Socas-Navarro:etal:2004}
{Socas-Navarro}, H., {Pillet}, V.~M., {Sobotka}, M., \& {V{\' a}zquez}, M.
  2004, \apj, 614, 448

\bibitem[{{Solanki}(2003)}]{Solanki:2003}
{Solanki}, S.~K. 2003, {\aap}r, 11, 153

\bibitem[{{Solanki} \& {Montavon}(1993)}]{Solanki:Montavon:1993}
{Solanki}, S.~K. \& {Montavon}, C.~A.~P. 1993, \aap, 275, 283

\bibitem[{{Spruit} \& {Scharmer}(2006)}]{Spruit:Scharmer:2006}
{Spruit}, H.~C. \& {Scharmer}, G.~B. 2006, \aap, 447, 343

\bibitem[{{Stanchfield} {et~al.}(1997){Stanchfield}, {Thomas}, \&
  {Lites}}]{Stanchfield:etal:1997}
{Stanchfield}, II, D.~C.~H., {Thomas}, J.~H., \& {Lites}, B.~W. 1997, \apj,
  477, 485

\bibitem[{{Thomas}(1988)}]{Thomas:1988}
{Thomas}, J.~H. 1988, \apj, 333, 407

\bibitem[{{Thomas}(2010)}]{Thomas:2010}
{Thomas}, J.~H. 2010, in Magnetic Coupling between the Interior and Atmosphere
  of the Sun, ed. {S.~S.~Hasan \& R.~J.~Rutten}, 229--242

\bibitem[{{Thomas} \& {Montesinos}(1993)}]{Thomas:Montesinos:1993}
{Thomas}, J.~H. \& {Montesinos}, B. 1993, \apj, 407, 398

\bibitem[{{Thomas} \& {Weiss}(1992)}]{Thomas:Weiss:1992}
{Thomas}, J.~H. \& {Weiss}, N.~O. 1992, in Sunspots: Theory and Observations,
  ed. J.~H. {Thomas} \& N.~O. {Weiss} (Kluwer, NATO ASI, C 375), 3--59

\bibitem[{{Thomas} \& {Weiss}(2004)}]{Thomas:Weiss:2004}
{Thomas}, J.~H. \& {Weiss}, N.~O. 2004, \araa, 42, 517

\bibitem[{{Thomas} \& {Weiss}(2008)}]{Thomas:Weiss:2008}
---. 2008, {Sunspots and Starspots}, ed. {Thomas, J.~H.~\& Weiss, N.~O.}
  (Cambridge University Press)

\bibitem[{{Tritschler} {et~al.}(2007){Tritschler}, {M{\"u}ller},
  {Schlichenmaier}, \& {Hagenaar}}]{Tritschler:etal:2007}
{Tritschler}, A., {M{\"u}ller}, D.~A.~N., {Schlichenmaier}, R., \& {Hagenaar},
  H.~J. 2007, \apjl, 671, L85

\bibitem[{{Vasil} \& {Brummell}(2009)}]{Vasil:Brummell:2009}
{Vasil}, G.~M. \& {Brummell}, N.~H. 2009, \apj, 690, 783

\bibitem[{{Wentzel}(1992)}]{Wentzel:1992}
{Wentzel}, D.~G. 1992, \apj, 388, 211

\bibitem[{{Westendorp Plaza} {et~al.}(2001){Westendorp Plaza}, {del Toro
  Iniesta}, {Ruiz Cobo}, \& {Mart{\'{\i}}nez Pillet}}]{Westendorp:etal:2001}
{Westendorp Plaza}, C., {del Toro Iniesta}, J.~C., {Ruiz Cobo}, B., \&
  {Mart{\'{\i}}nez Pillet}, V. 2001, \apj, 547, 1148

\bibitem[{{Zakharov} {et~al.}(2008){Zakharov}, {Hirzberger}, {Riethm{\"u}ller},
  {Solanki}, \& {Kobel}}]{Zakharov:etal:2008}
{Zakharov}, V., {Hirzberger}, J., {Riethm{\"u}ller}, T.~L., {Solanki}, S.~K.,
  \& {Kobel}, P. 2008, \aap, 488, L17

\bibitem[{{Zhao} {et~al.}(2001){Zhao}, {Kosovichev}, \&
  {Duvall}}]{Zhao:etal:2001}
{Zhao}, J., {Kosovichev}, A.~G., \& {Duvall}, T.~L. 2001, \apj, 557, 384

\bibitem[{{Zhao} {et~al.}(2010){Zhao}, {Kosovichev}, \&
  {Sekii}}]{Zhao:etal:2010}
{Zhao}, J., {Kosovichev}, A.~G., \& {Sekii}, T. 2010, \apj, 708, 304

\end{thebibliography}

\end{document}